\documentclass[lettersize,journal]{IEEEtran}
\usepackage{amsmath,amsfonts}
\usepackage{algorithmic}
\usepackage{algorithm}
\usepackage{array}
\newcommand{\eg}{{\it e.g.}}
\newcommand{\ie}{{\it i.e.}}
\usepackage[caption=false,font=normalsize,labelfont=sf,textfont=sf]{subfig}
\usepackage{textcomp}
\usepackage{stfloats}
\usepackage{url}
\usepackage{verbatim}
\usepackage{graphicx}
\usepackage{multirow}
\usepackage{cite}
\usepackage{booktabs}
\usepackage{xcolor}
\usepackage{color}
\newcommand{\mrh}{\textcolor[rgb]{0.0, 0.0, 0.0}}

\begin{document}

\title{Semantic Deep Hiding for \\Robust Unlearnable Examples
% Towards Robust Unlearnable Examples \\via Deep Hiding
}

\author{Ruohan Meng%~\IEEEmembership{Member,~IEEE}
, Chenyu Yi*, Yi Yu, Siyuan Yang, Bingquan Shen, and Alex C. Kot,~\IEEEmembership{Life Fellow,~IEEE}
        % <-this % stops a space
%\thanks{This paper was produced by the IEEE Publication Technology Group. They are in Piscataway, NJ.}% <-this % stops a space
\thanks{
{Ruohan Meng, Chenyu Yi, Yi Yu, Siyuan Yang, and Alex C. Kot are with the School of Electrical and Electronic Engineering, Rapid-Rich Object Search (ROSE) Laboratory, Nanyang Technological University, 639798, Singapore. (e-mail: ruohan.meng@ntu.edu.sg; cyyi@ntu.edu.sg; yuyi0010@e.ntu.edu.sg; siyuan.yang@ntu.edu.sg; eackot@ntu.edu.sg)}

{Bingquan Shen is with the DSO National Laboratories, Singapore. (e-mail: sbingqua@dso.org.sg)}

Correspondence author: Chenyu Yi.

This research work was carried out at the Rapid-Rich Object Search (ROSE) Lab, the NTU-PKU Joint Research Institute (sponsored by the Ng Teng Fong Charitable Foundation), Nanyang Technological University, Singapore. The research is supported by the DSO National Laboratories, under project agreement No. DSOCL22332.
}}

\markboth{Journal of \LaTeX\ Class Files,~Vol.~14, No.~8, August~2021}%
{Shell \MakeLowercase{\textit{et al.}}: A Sample Article Using IEEEtran.cls for IEEE Journals}

\IEEEpubid{0000--0000/00\$00.00~\copyright~2021 IEEE}

\maketitle

\begin{abstract}
Ensuring data privacy and protection has become paramount in the era of deep learning. 
Unlearnable examples are proposed to mislead the deep learning models and prevent data from unauthorized exploration by adding small perturbations to data.
{However, such perturbations (\eg, noise, texture, color change) predominantly impact low-level features, making them vulnerable to common countermeasures.}
In contrast, semantic images with intricate shapes have a wealth of high-level features, making them more resilient to countermeasures and potential for producing robust unlearnable examples.
In this paper, we propose a Deep Hiding (DH) scheme that adaptively hides semantic images enriched with high-level features. We employ an Invertible Neural Network (INN) to invisibly integrate predefined images, inherently hiding them with deceptive perturbations.
To enhance data unlearnability, we introduce a Latent Feature Concentration module, designed to work with the INN, regularizing the intra-class variance of these perturbations.
To further boost the robustness of unlearnable examples, we design a Semantic Images Generation module that produces hidden semantic images. 
By utilizing similar semantic information, this module generates similar semantic images for samples within the same classes, thereby enlarging the inter-class distance and narrowing the intra-class distance.

\mrh{
Extensive experiments on CIFAR-10, CIFAR-100, and an ImageNet subset, against 18 countermeasures, reveal that our proposed method exhibits
outstanding robustness for unlearnable examples, demonstrating its efficacy in preventing unauthorized data exploitation.
}

\end{abstract}


\section{Introduction}
\begin{figure}[t]
\begin{center}
\includegraphics[width=0.49\textwidth]{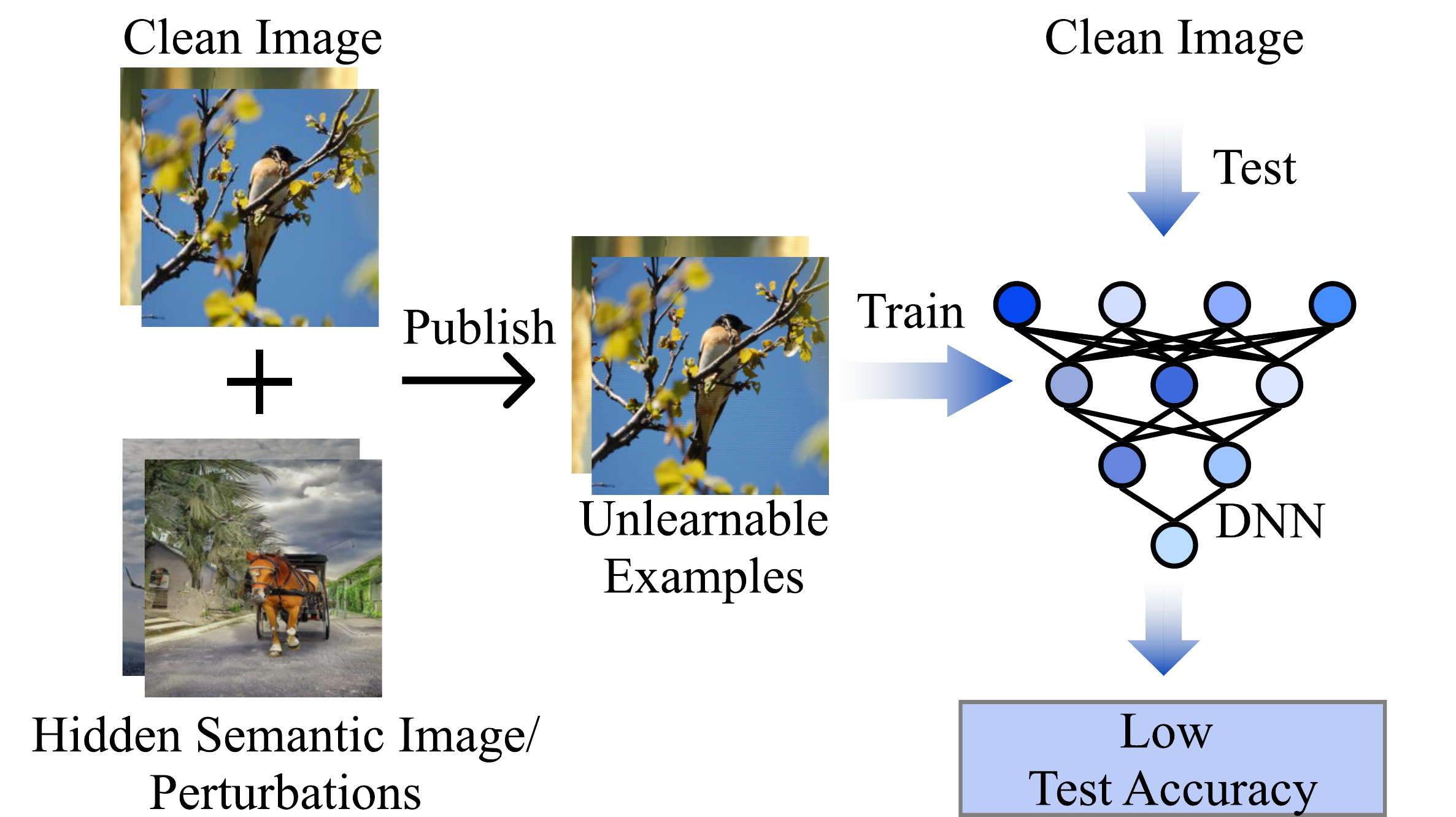}
\end{center}
\vspace{-5mm}
\caption{
An illustration depicts the concept of unlearnable examples, where semantic images are cleverly embedded within clean images to create these unlearnable examples. When Deep Neural Networks (DNNs) are trained on these examples, they fail to learn the meaningful features of the clean images due to the embedded semantic perturbations. Consequently, the accuracy of the DNN models on clean data becomes significantly and unpredictably poor.
}
 \vspace{-3mm}
\label{Fig:illustration}
\end{figure}

\IEEEPARstart{T}{he} rapid growth of deep learning is largely attributed to the vast amounts of “free" data available on the internet. However, a significant portion of these datasets might encompass personal information obtained without clear authorization~\cite{intro1,intro2}. Such practices have heightened societal concerns regarding the potential misuse of individual data, particularly when leveraged to develop commercial or potentially malicious models absent the owner's consent~\cite{privacy}. To address these concerns, the concept of unlearnable examples~\cite{TC, DC, EM, HYPO, TAP,lin2024safeguarding} was introduced, which aims to prevent a deep learning model's ability to discern meaningful features from genuine patterns by introducing minor perturbations to clean images, as shown in Fig.~\ref{Fig:illustration}.

{When we deploy unlearnable examples to protect unauthorized data in the real world, their strong robustness against different countermeasures plays a critical role~\cite{radiya2021data}}. {Existing methods~\cite{REM, ADVIN, EntF} mainly focus on improving their robustness against adversarial training~\cite{madry2017towards,yu2022towards,xia2024mitigating}}, since the unlearnable examples like error-minimization~\cite{EM} or targeted adversarial poison~\cite{TAP} show vulnerability to adversarial training~\cite{madry2017towards}. However, the general robustness of unlearnable examples against various countermeasures (\eg, data augmentations, data preprocessing) has been ignored. For example,  Image Shortcut Squeezing~\cite{ISS} reveals simple JPEG compression and grayscale transformation can significantly impact the effectiveness of most existing unlearnable examples methods; OPS~\cite{OPS} demonstrates strong adversarial robustness, while it is extremely fragile to widely used operations including cutout and median filtering.
\mrh{Consequently, we introduce a Deep Hiding scheme, termed DH, designed to generate robust unlearnable examples by adaptively hiding semantic images enriched with high-level features as poisons, thereby providing fortified protection against unauthorized data exploitation.}
{Several studies~\cite{shape, high1, high2, zhou2023generative, high3} indicate that the natural image with semantic information (\eg, intricate shapes) is robust against data augmentations, data preprocessing, and adversarial training. } \IEEEpubidadjcol
Additionally, the existing image hiding techniques~\cite{baluja2017hiding, yu2020attention, Hinet,zhang2019invisible, tang2020automatic, pan2021seek, cui2024meta, hu2023invisible} support adaptively hiding one image within another. Among them, the Invertible Neural Networks (INNs)~\cite{Hinet, deepmih, INN-based, meng2022traceable, Traceble} are notable for their outstanding capability to render images virtually invisible.

Specifically, our proposed method employs an INN model to invisibly and adaptively hide semantic images, endowed with rich high-level attributes, into clean images, generating deceptive perturbations.
To enhance the effectiveness of unlearnable examples, we introduce the Latent Feature Concentration module (LFC) to limit intra-class variance by regularizing the latent feature distance of the perturbations. Additionally, we design a Semantic Images Generation module to produce hidden semantic images, by controlling the semantic features (\ie, shapes, {edges}) during the generation process.
Capitalizing on similar semantic information, this module generates analogous semantic images for samples within identical categories.
These modules increase the inter-class separation and minimize the intra-class variance, enhancing the robustness of unlearnable examples.

\mrh{In our designed scheme, the deep learning model prioritizes the semantic features of hidden images over those of genuine patterns due to the semantic nature of the hidden features.}
Additionally, semantic images with complex shapes possess rich high-level attributes that exhibit greater resistance to data countermeasures.
In the experiments, we implemented two settings of hidden semantic images: class-wise and sample-wise, aligning them to a single class to strike a balance between efficiency and exposure risk.
\mrh{Extensive experiments conducted on CIFAR-10, CIFAR-100, and a subset of ImageNet demonstrate that our method outperforms the vast majority of countermeasures.}
{Across common countermeasures, the ResNet-18~\cite{resnet18} models trained on the perturbed CIFAR-10, CIFAR-100 and ImageNet subset have average test accuracy of 33.01\%, 18.95\% and 11.39\% respectively, compared to the best performance of 39.20\%, 23.17\% and 22.76\% by the other unlearnable examples techniques.} Our contributions can be summarized as:
\begin{itemize}    
    \item \mrh{We conceptualize the generation process of unlearnable examples as an image-hiding challenge. \mrh{To address this, we first introduce a Deep Hiding scheme that invisibly and adaptively hides semantic images, enriched with high-level attributes, into clean images using an INN model}.}
    \item {We propose the Latent Feature Concentration module, designed to regularize the intra-class variance of perturbations, enhancing the effectiveness of unlearnable examples.
    Moreover, we design the Semantic Images Generation module to generate hidden semantic images by maintaining semantic feature consistency within a single class, aiming to amplify the robustness of unlearnable examples.}
        \item {
       \mrh{
       Extensive experiments conducted on CIFAR-10, CIFAR-100, and a subset of ImageNet demonstrate that our method of deep hiding semantic features as poisons effectively exhibits outstanding robustness against most countermeasures. 
       Both the average and maximum test accuracies consistently show superior performance for our method, highlighting its efficacy in preventing unauthorized data exploitation.
       }
        }
\end{itemize}

\section{Related Work}
\label{gen_inst}

\subsection{Unlearnable examples} {To safeguard data from unauthorized scraping, there is an emerging research emphasis on techniques to render data ``unlearnable" for machine learning models.
Considering the surrogate models utilized in training, denoted as surrogate-dependent models, Targeted Adversarial Poisoning (TAP)~\cite{TAP} employs adversarial examples as a more effective form of data poisoning, aiming to ensure that models trained on adversarially perturbed data fail to identify even their original counterparts.}
Building on this, Error-Minimizing (EM)~\cite{EM} introduces the concept of “unlearnable examples” and employs “error-minimizing noise” through a bi-level optimization process to make data unlearnable.
However, this approach is not robust against adversarial training~\cite{madry2017towards}. 
To address this limitation, Robust Error-Minimizing (REM)~\cite{REM} introduces a robust error-minimizing noise by incorporating adversarial training and the expectation over transformation~\cite{EOT} technique. 
Further enhancing the utility of unlearnable examples, ADVersarially Inducing Noise (ADVIN)~\cite{ADVIN} and Entangled Features (EntF)~\cite{EntF} propose similar methods to enhance the robustness of adversarial training.
On another front, Transferable Unlearnbale Examples (TUE)~\cite{TUE} proposes the classwise separability discriminant to improve their transferability across different training settings and datasets. 
{Unlearnable Clusters
(UCs)~\cite{zhang2023unlearnable} introduces label-agnostic unlearnable examples with cluster-wise
perturbations without knowing the label information. TUE and UCs can prevent unsupervised
exploitation against contrastive learning to a certain extent.}
However, the generated perturbation derived from gradient learning strongly requires knowledge from the surrogate model.
In contrast, Autoregressive (AR)~\cite{AR} introduces a surrogate-free methodology, proposing AR perturbations that remain independent of both data and models. 
Besides, Linear separable Synthetic Perturbations (LSP)~\cite{LSP} investigates the underlying mechanisms of availability attacks, identifying that the perturbations serve as “shortcuts” for learning objectives, and introducing synthetic shortcuts by sampling from Gaussian distributions.
Another novel approach is One Pixel Shortcut (OPS)~\cite{OPS}, 
a single pixel in each image results in significant degradation of model accuracy.

\subsection{Robustness} 
In different stages of deep learning model training, the unlearnable examples will encounter various data manipulations. For example, in the data pre-processing, the unlearnable examples could be corrupted by noise and compression effects; before we feed them into the model for training, we usually apply data augmentations. However, certain data manipulations can diminish the efficacy of the added perturbation~\cite{ISS, zhang2023apmsa, xu2022robust, zhu2023information, lee2023robust}. It is crucial to make the unlearnable examples generally robust against these data manipulations in real-world applications.

To evaluate the robustness of these generated unlearnable examples, Image Shortcut Squeezing (ISS)~\cite{ISS} utilizes fundamental countermeasures based on image compression techniques, such as grayscale transformation, JPEG compression, and bit-depth reduction (BDR), to counteract the effects of perturbations. 
In addition, techniques such as Gaussian blur, random cropping and flipping, Cutout~\cite{cutout}, Cutmix~\cite{cutmix}, and Mixup~\cite{mixup} are employed to assess the robustness of unlearnable examples. Additionally, as referenced in the unlearnable examples part, adversarial training (AT) stands as a pivotal method to bolster the resilience of crafted unlearnable examples.
{More contemporarily, UEraser~\cite{qin2023learning} proposes a combination of effective data augmentation policies and loss-maximizing adversarial augmentations to attack the existing unlearnable examples.} 
AVATAR~\cite{AVATAR} extends the methodology outlined in DiffPure~\cite{nie2022diffusion}, using diffusion models to counteract intentional perturbations while preserving the essential semantics of training images. 
{And Orthogonal Projection (OP)~\cite{sandoval2024can} presents a new attack that allows learning from unlearnable datasets perturbed by classwise, linearly separable perturbations.}
D-VAE~\cite{yu2024purify} adopts rate-constrained variational autoencoders with perturbations disentanglement to purify the unlearnable datasets.

\subsection{Image hiding}
{As deep learning continues to evolve, researchers are exploring methods to seamlessly embed whole images within other images using deep neural networks such as encoder-decoder~\cite{baluja2017hiding, yu2020attention}, GAN-based~\cite{zhang2019invisible}, and reinforcement learning-driven models~\cite{pan2021seek} to subtly embed one or multiple images within a container image. 
Leveraging the inverse property of INN for image-to-image tasks~\cite{zhao2021invertible, huang2022winnet}, HiNet~\cite{jing2021hinet} and DeepMIH~\cite{guan2022deepmih} employ DWT to decompose the input image, and constrain the hiding to implementation in high-frequency sub-bands for invisible hiding. 
Similarly, iSCMIS~\cite{li2023iscmis}, ISN~\cite{lu2021large}, and RIIS~\cite{xu2022robust} hide data by using the inverse property, 
{with some models even simulating data transformations to enhance the robust retrieval of hidden data.}
% and some simulate the data transformations for robust extraction of hidden information. 
According to the above methods, INN shows good potential to achieve image hiding. Note that the robustness in image hiding fields means the extraction performance of generated images undergoing some attacks. However, the robustness of unlearnable examples should be focused on unlearnable stability. 
In the backdoor and adversarial attack fields, image hiding schemes have notably contributed. 
Specifically, the Backdoor Injection attack~\cite{zhong2020backdoor} utilizes INN to generate robust and subtle adversarial examples, yielding adversarial images that are both less noticeable and more resilient compared to traditional techniques. 
Poison Ink~\cite{zhang2022poison} exploits image structures, employing image hiding schemes to incorporate trigger patterns, thereby ensuring stealthy and robust backdoor attacks. 
AdvINN~\cite{chen2023imperceptible} utilizes INN to generate robust and subtle adversarial examples, yielding adversarial images that are both less noticeable and more resilient compared to traditional techniques.
Such strategies underscore the significant potential of deep image hiding in bolstering the effectiveness of unlearnable examples.

\section{Proposed Method} 
\label{headings}
\subsection{Definition}
\subsubsection{Recalling unlearnable examples}
Following the existing unlearnable research~\cite{EM, REM, TAP, AR, LSP}, we focus on the image classification task in this work.
Given a clean dataset $\mathcal{D}_c=\{\left(\boldsymbol{x}_i, y_i\right)\}_{i=1}^n$ with $n$ training samples, where $\boldsymbol{x} \in \mathcal{X} \subset \mathbb{R}^d$ represents the images and $y \in \mathcal{Y}=\{1, \cdots, K\}$ denotes its corresponding labels.
We assume an unauthorized party will use a classifier given as $f_\theta: \mathcal{X} \rightarrow \mathcal{Y}$ where $\theta \in \Theta$ is the classifier parameters.
To safeguard the images from unauthorized training, rather than publishing $\mathcal{D}_c$, existing methods~\cite{EM, REM} introduce perturbation to clean images, generating an unlearnable dataset as:
\begin{equation}
    \mathcal{D}_u=\left\{\left(\boldsymbol{x}_i+\boldsymbol{\delta}_i, y_i\right)\right\}_{i=1}^n,
\end{equation}
where $\boldsymbol{\delta}_i \in \Delta_{\mathcal{D}} \subset \mathbb{R}^d$ and $\Delta_{\mathcal{D}}$ is the perturbation set for $\mathcal{D}_c$.
{The objective of unlearnability is to ensure that a classifier $f_\theta$ trained on $\mathcal{D}_u$ exhibits poor performance on test datasets.}

\subsubsection{Proposed unlearnable examples}
Current approaches typically generate perturbations either through gradient-based training with a surrogate model or by sampling noise from a predefined distribution in model-free manners.
{These perturbations lack distinguished semantic high-level features and redundancy, making it challenging to generalize robustness against various countermeasures. }
{It is worth noting that while the perturbations generated in REM and EntF exhibit certain semantic patterns, these patterns are tailored to resist AT and are closely aligned with the clean image's features. It compromises their effectiveness against other countermeasures.}
{Conversely, we propose an adaptive method for embedding a distinguished semantic image $\boldsymbol{h}_i$ characterized by rich high-level features and completely uncorrelated with the features of the clean image, within a clean image to generate unlearnable examples.}
Thus, the generated unlearnable dataset is defined as: 
\begin{equation}
    \mathcal{D}_u=\left\{\left(\mathcal{F} (\boldsymbol{x}_i, \boldsymbol{h}_i), y_i\right)\right\}_{i=1}^n,
\end{equation}
where $\mathcal{F}(\cdot, \cdot)$ represents our hiding model. 
We adaptively hide predefined semantic images into clean datasets $\mathcal{D}_{c}$ rather than arbitrary perturbation, inherently introducing deceptive perturbations to mislead classifier $f_\theta$, thereby enhancing the effectiveness of unlearnable examples.
% robustness

\begin{figure*}[t]
\begin{center}
%\framebox[4.0in]{$\;$}
\includegraphics[width=0.97\textwidth]{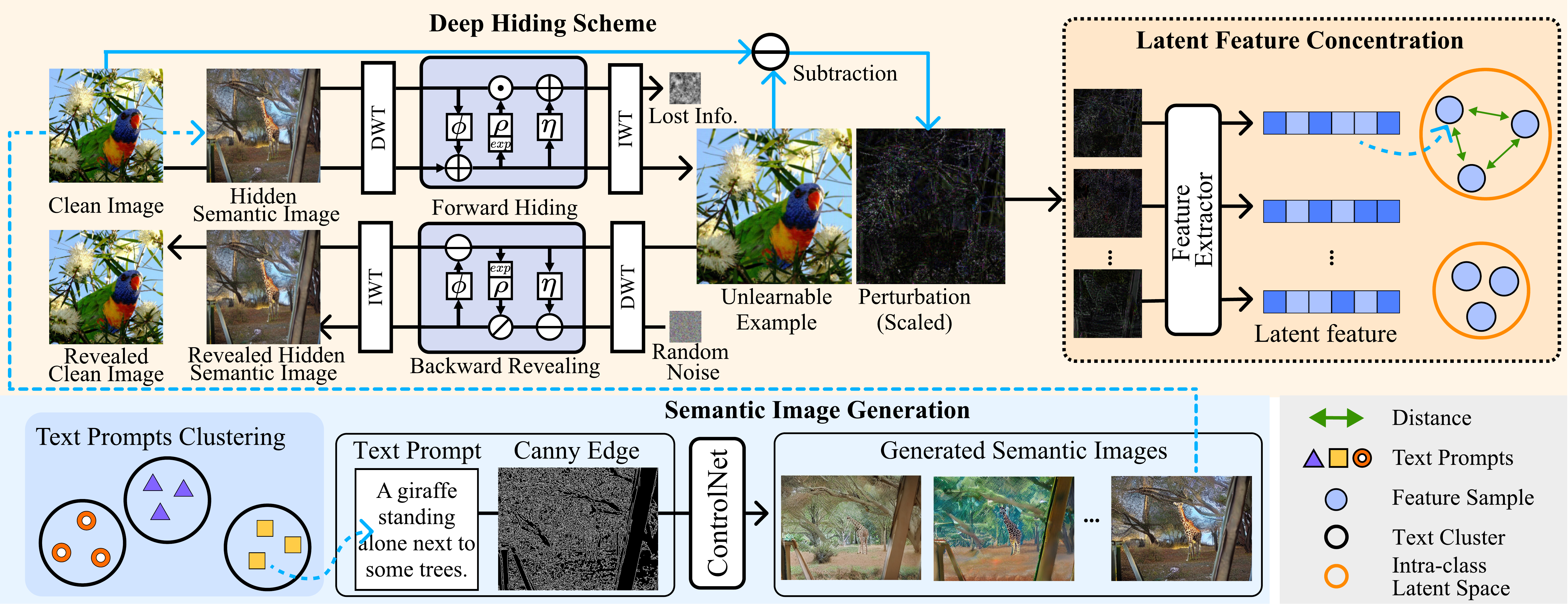}
%\fbox{\rule[-.5cm]{0cm}{4cm} \rule[-.5cm]{4cm}{0cm}}
\end{center}
\vspace{-4mm}
\caption{Overall pipeline of the proposed method. 
{A generative model is employed to generate the hidden semantic images. These generated images are then hidden within clean images using a Deep Hiding model. The Latent Feature Concentration module is designed to constrain the intra-class variance by regularizing the latent feature distance of perturbations.  
}
% \vspace{-4mm}
}
\label{Fig:Overview}
\end{figure*}
\subsection{Deep hiding scheme for robust unlearnable examples}
To generate a resilient unlearnable dataset $\mathcal{D}_{u}$, we introduce the Deep Hiding scheme. 
This framework incorporates the image hiding model, which integrates the specially-designed Latent Feature Concentration module, and the Semantic Images Generation module. 
Fig.~\ref{Fig:Overview} illustrates the overview of the proposed Deep Hiding scheme.

\subsubsection{Deep hiding model}
\label{dh}
Inspired by the image hiding methods~\cite{baluja2017hiding, yu2020attention, Hinet,zhang2019invisible, pan2021seek}, we employ the INN-based hiding model{, HiNet~\cite{jing2021hinet}, as our framework, leveraging its superior {generation} performance.}
{HiNet employs $N$ affine coupling blocks to form two invertible processes, forward hiding and backward revealing, where the hiding process enables inherently embedding predefined semantic images into clean images, as illustrated in Fig.~\ref{Fig:Overview}. 
To {facilitate} invisible deep hiding, Discrete Wavelet Transform (DWT) $\mathcal{T}(\cdot)$ is applied to decompose}
%to the inputs for decomposition to disentangle 
{the input clean image $\boldsymbol{x}_{c}$, and hidden semantic image $\boldsymbol{x}_{h}$
into low and high-frequency sub-bands.
We 
{denote} the sub-bands features of $\boldsymbol{x}_{c}$ and $\boldsymbol{x}_{h}$ as $\boldsymbol{z}_{c} = \mathcal{T}(x_{c})$ and $\boldsymbol{z}_{h} = \mathcal{T}(x_{h})$, respectively.}
{Let $\boldsymbol{z}_{c}^i$ and {$\boldsymbol{z}_{h}^i$} 
%are denoted as 
{be} the input features of the $i^{th}$ affine coupling block, %. 
the forward hiding process of 
%the $i^{th}$ affine coupling 
{this} block can be expressed as{:}}%,
{\iffalse
\begin{equation}
\begin{aligned} 
\boldsymbol{z}_{c}^i=&\boldsymbol{z}_{c}^{i-1}+\phi\left(\boldsymbol{z}_{h}^{i-1}\right),\\ \boldsymbol{x}_{h}^i=&\boldsymbol{z}_{h}^{i-1} \odot \exp \left(\alpha\left(\rho\left(\boldsymbol{z}_{c}^i\right)\right)\right)+\eta\left(\boldsymbol{z}_{c}^i\right),
\end{aligned}
\end{equation}
\fi
\begin{equation}
\begin{aligned} 
\boldsymbol{z}_{c}^i=\boldsymbol{z}_{c}^{i-1}+\phi\left(\boldsymbol{z}_{h}^{i-1}\right),
\end{aligned}
\end{equation}
\begin{equation}
\begin{aligned} 
\boldsymbol{z}_{h}^i=\boldsymbol{z}_{h}^{i-1} \odot \exp \left(\alpha \cdot \rho\left(\boldsymbol{z}_{c}^i\right)\right)+\eta\left(\boldsymbol{z}_{c}^i\right),
\end{aligned}
\end{equation}
where $\phi(\cdot)$, $\rho(\cdot)$, and $\eta(\cdot)$ are 
{three sub-modules, sharing the same network structure but with different weights,}
%sub-modules with convolution operations, 
$\exp (\cdot)$ is the Exponential function, %and
$\odot$ is the Hadamard product operation, {and} $\alpha$ controls the weight of exponential operation.
Given the output {features $\boldsymbol{z}_{c}^{N}$} of {total} $N^{th}$ affine coupling block, the unlearnable examples $\boldsymbol{x}_{ue} = \mathcal{T}^{-1}(\boldsymbol{z}_{c}^{N})$ %is
{are} generated %from the $N^{th}$ features via IWT.
{by Inverse DWT (IDWT).}}

{To ensure the success of the image-hiding procedure, in the backward revealing process, the obtained unlearnable examples} %image is randomly sampled latent noises
are first decomposed by DWT and then together with the randomly sampled latent noises $r$ 
feed into the HiNet, 
%served as the input into the INN, 
resulting in the revealed clean image 
{$\boldsymbol{x}^{\prime}_c=\mathcal{T}^{-1}(\boldsymbol{z}_{c}^{1})$ and revealed hidden semantic image $\boldsymbol{x}^{\prime}_h = \mathcal{T}^{-1}(r^{1})$ by subsequent IDWT}.
{Such $\boldsymbol{z}_{c}^{1}$ and $r^1$ can be obtained by:  }
\begin{equation}
\begin{aligned} 
 \boldsymbol{z}_{c}^{i-1}=\left(\boldsymbol{z}_{c}^i-\eta\left(\boldsymbol{z}_{c}^i\right)\right) \odot \exp \left(-\alpha \cdot \rho\left(\boldsymbol{z}_{c}^i\right)\right), 
\end{aligned}
\end{equation}
\begin{equation}
\begin{aligned} 
 {r}^{i-1}={r}^i-\phi\left({r}^{i-1}\right).
\end{aligned}
\end{equation}

Our primary objective is to generate invisible unlearnable examples. To ensure this, motivated by \cite{yu2023backdoor}, we restrict them to a specific radius $\epsilon$ of perturbation, characterized by the hiding loss as:
\begin{equation}
\begin{aligned} 
\mathcal{L}_{\text{hide}}\left(\boldsymbol{x}_{ue}, \boldsymbol{x}_{c}\right) = \max\left(\text{MSE}(\boldsymbol{x}_{ue}, \boldsymbol{x}_{c}),\epsilon^2\right).
\end{aligned}
\end{equation}
{For consistency and fairness, we adopt the same radius $\epsilon=8/255$ as utilized in existing unlearnable examples methodologies~\cite{EM, REM, TAP}.}

In addition, we adapt the loss functions from HiNet~\cite{Hinet} to concurrently ensure optimal image hiding performance. Consequently, the total loss for the Deep Hiding module is represented as follows:
\iffalse
\begin{equation}\label{eqn:baseloss}
\begin{aligned} 
\!{\mathcal{L}_{\text{DH}} & \!= \!\mathcal{L}_{\text{hide}}\left(\boldsymbol{x}_{ue}, \boldsymbol{x}_{c}\right) \!+\! \omega_1 \!\cdot \! \mathcal{L}_{\text{freq}}\left(\mathcal{H}\left(\boldsymbol{x}_{ue}\right)_{L L}, \mathcal{H}\left(\boldsymbol{x}_{c}\right)_{L L}\right)\\
&+\omega_2\cdot\mathcal{L}_{\text{reveal}}\left(\boldsymbol{x}^{\prime}_{h}, \boldsymbol{x}_{h}\right),}
\end{aligned}
\end{equation}
\fi
\begin{equation}\label{eqn:baseloss}
\begin{aligned}
\!\!\!\mathcal{L}_{\text{DH}} \!= & \mathcal{L}_{\text{hide}}\left(\boldsymbol{x}_{ue}, \boldsymbol{x}_{c}\right) \!+ \omega_1 \!\cdot \mathcal{L}_{\text{freq}}\left(\mathcal{H}\left(\boldsymbol{x}_{ue}\right)_{LL}, \mathcal{H}\left(\boldsymbol{x}_{c}\right)_{LL}\right) \\
& + \omega_2 \cdot \mathcal{L}_{\text{reveal}}\left(\boldsymbol{x}^{\prime}_{h}, \boldsymbol{x}_{h}\right).
\end{aligned}
\end{equation}

As verified by~\cite{Hinet, deepmih}, 
information hidden in high-frequency components is less detectable than in low-frequency ones.
To optimize the anti-detection and invisibility of unlearnable examples, it's crucial to maintain the low-frequency sub-bands to closely resemble those of clean images. 
$\mathcal{L}_{\text{freq}}$ measures the $L_2$ distance between the low-frequency sub-bands of clean images and unlearnable examples, further bolstering the stealthiness. {$\mathcal{H}(\cdot)_{LL}$ is the function of extracting low-frequency sub-bands after wavelet decomposition.}
Additionally, $\mathcal{L}_{\text{reveal}}\left(\boldsymbol{x}^{\prime}_{h}, \boldsymbol{x}_{h}\right)$ measures the $L_2$ 
distance between revealed hidden images $\boldsymbol{x}^{\prime}_{h}$ and hidden semantic images $\boldsymbol{x}_{h}$. {It's important to note that the revealing process is crucial for reserving the semantic information in image hiding.}
\subsubsection{Latent feature concentration}
Although the deep hiding model effectively embeds high-level features from predefined semantic images into clean images, delivering outstanding unlearnability (see Section~\ref{sec:overall_performance}), the adaptive hiding process still results in latent features of perturbations with non-uniform intra-class distribution.
A compact distribution of these latent features could significantly mislead the learning trajectory of DNNs, by offering a distinct learning shortcut across similar intra-class images. 
To address this, we introduce the Latent Feature Concentration module, specifically 
designed to regularize the intra-class variance of perturbations, further boosting the effectiveness of unlearnable examples.
{The perturbation represents the variation between the generated unlearnable example and its corresponding clean image, defined as: 
\begin{equation}
\boldsymbol{x}_{pm} = \boldsymbol{x}_{ue} - \boldsymbol{x}_{c}.
\end{equation}
We utilize a pre-trained ResNet-18~\cite{resnet18} as the feature extractor, denoted by $\mathcal{G}(\cdot)$.
The latent features are extracted from the output final convolution layer. 
Our objective is to minimize the variation between latent features derived from the perturbation maps for images within the same class.
Consequently, the concentration loss $\mathcal{L}_{\text{conc}}$ is represented as:}
\begin{equation}
\begin{aligned} 
\mathcal{L}_{\text{conc}} = \sum_{i,j,y^{i}=y^{j}} dis\left( \mathcal{G} (\boldsymbol{x}^{i}_{pm}), \mathcal{G}(\boldsymbol{x}^{j}_{pm})\right) ,
\end{aligned}
\end{equation}
where $dis(\cdot, \cdot)$ denotes the cosine distance between the two flattened latent features, and $y$ represents the label.
Thus, the total loss of our proposed method is described as:
\begin{equation}
\begin{aligned} 
\mathcal{L}_{\text{total}}= \mathcal{L}_{\text{DH}} + \omega_3\cdot\mathcal{L}_{\text{conc}}.
\end{aligned}
\end{equation}

\subsubsection{Semantic images generation}
\label{image_gen}
Though the deep hiding model can embed human-imperceptible perturbations, it can not ensure efficacy when the hidden images are randomly picked. Consequently, we introduce a generative method specifically designed for controlled hidden semantic image generation, aiming to achieve desired intra-class and inter-class distributions; that is, a distinct inter-class distance complemented by a minimal intra-class distance. As shown in Fig.~\ref{Fig:Overview}, we use pre-trained generative models, \ie, Stable Diffusion~\cite{ldm} and ControlNet~\cite{controlnet}, to generate hidden semantic images by controlling both text prompts and canny edge maps. These text prompts, sourced from~\cite{prompt}, characterize images from the COCO datasets~\cite{coco}.
The canny edge maps are derived by applying the canny filter to the corresponding images.

To maximize the intra-class distance among hidden semantic images, we choose text prompts that exhibit the greatest variation from the rest.
We first cluster all text prompts using K-Means~\cite{kmeans} based on their semantic features via CLIP text extractor~\cite{clip}. Subsequently, we identify the top-$k$ distinct clusters, where $k$ represents the number of classes in the targeted dataset $\mathcal{D}_c$.
In each of these clusters, we choose the text prompt nearest to the cluster center, which represents a unique semantic feature.
To minimize the intra-class distance among hidden semantic images, we ensure their consistency in high-level features by controlling the key image attributes, \ie, shapes. 
Consequently, we obtain a canny edge map of the text-corresponding image, which acts as the condition for ControlNet~\cite{zhang2023adding}.
Then, we use the Stable Diffusion~\cite{rombach2022high} model and ControlNet (SD+C) to generate semantic images as the hidden semantic input $\boldsymbol{x}_{h}$ for our DH scheme. With these specifically generated hidden semantic images, our deep hiding model can guarantee the general robustness of the unlearnable examples.

\begin{figure*}[t]
\begin{center}
\includegraphics[width=1\textwidth]{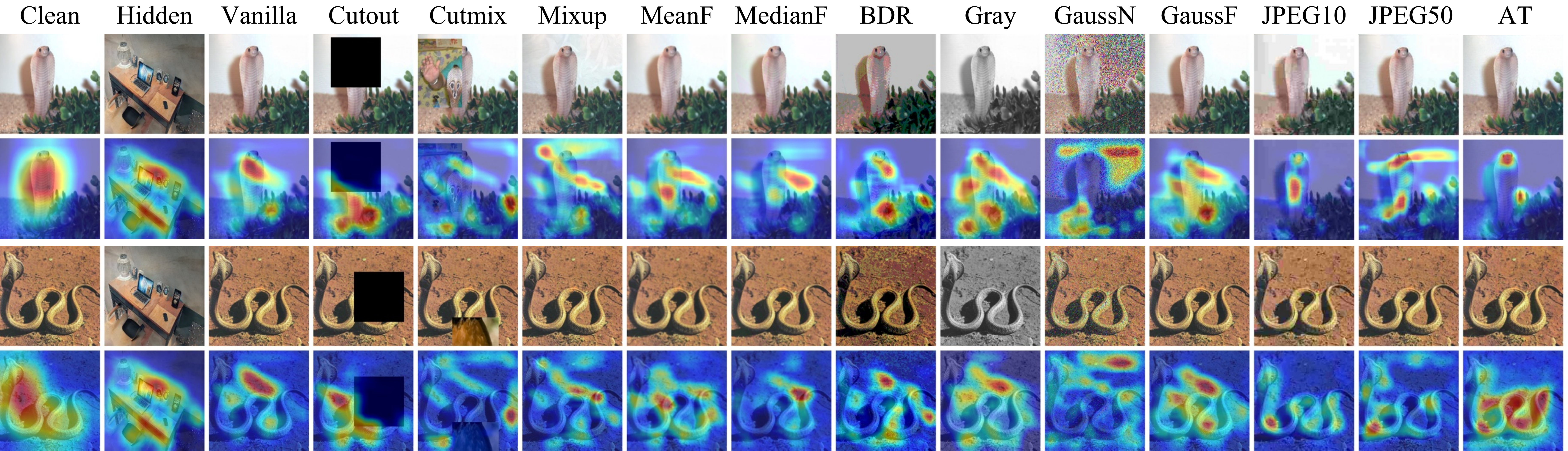}
\end{center}
\vspace{-4mm}
\caption{Grad-CAM visualization of unlearnable examples derived from the ImageNet subset under different countermeasures. 
Note that red regions typically indicate the areas the model paid the most attention to, while  {Blue} regions colors indicate less attention.}
\label{Fig:Grad}
\end{figure*}

\begin{figure}
\centering
\includegraphics[width=0.48\textwidth]{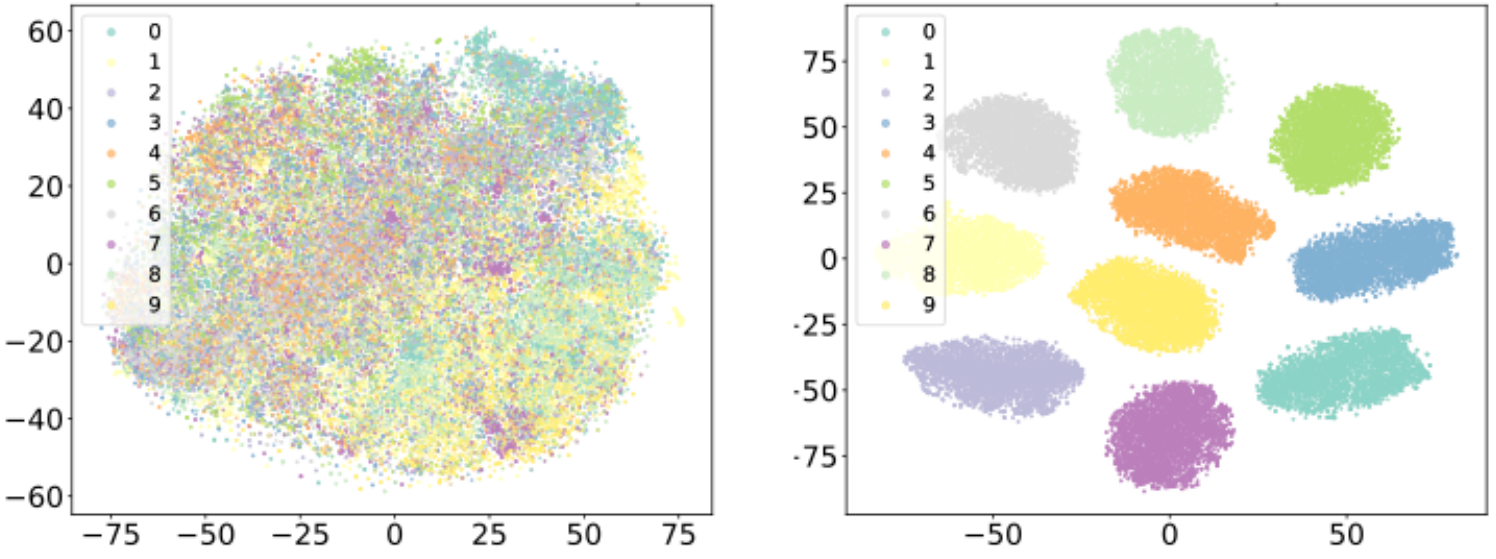}
\vspace{-4mm}
\caption{The t-SNE visualization of the models' feature representations on the clean samples (left) and the perturbation generated by our DH scheme (right) on CIFAR-10. }
\label{Fig:tsne}
\end{figure}
\subsection{Properties of deep hiding scheme}
DNNs are capable of learning complex features for image understanding.
However, they are inclined to overfit to the patterns that are much \textit{easier} to learn~\cite{geirhos2020shortcut}, in alignment with the ``Principle of Least Effort"~\cite{zipf2016human}.
With this phenomenon, many unlearnable examples are proposed to protect the data from being learned by DNNs.
Consequently, DNNs predominantly focus on misleading perturbations rather than the intended solutions. 
Our Deep Hiding scheme exploits the same principle. 
In our proposed scheme, clean images within a given class are embedded with similar hidden semantic images that share the same global shape but differ in their local textures. 
Compared to the complex features in the original images, the embedded similar semantic information is much \textit{easier} to be learned by DNNs. 
The visual representation in Fig.~\ref{Fig:tsne} demonstrates that the perturbations generated by our scheme exhibit clear separability, marked by straightforward decision boundaries.
Besides, We utilize Grad-CAM~\cite{grad} to visualize the attention of DNNs in Fig.~\ref{Fig:Grad}.
It is obvious that the attention is redirected toward the desk (the hidden semantic image) rather than the snake (the clean image) during classification. 
While DNNs can take non-semantic features as ``shortcuts'' for more effortless learning, these features are vulnerable to simple data augmentations and data processing.
Different from the existing unlearnable examples methods, we incorporate natural images as hidden semantic images to generate perturbations. 
These perturbations, enriched with deep high-level semantic attributes, exhibit robustness against diverse countermeasures.
As illustrated in Fig.~\ref{Fig:Grad}, the hidden semantic information can mislead the DNNs to similar key regions after most countermeasures. These findings affirm the efficacy and resilience of using natural images as hidden semantic information when faced with various countermeasures.

\section{Experiments}
\label{others}
\subsection{Experimental setups}
\subsubsection{Datasets and models} We use three image classification datasets: CIFAR-10~\cite{cifar10}, CIFAR-100~\cite{cifar10}, and 100-class subset of ImageNet~\cite{imagenet} in our experiments. 
{We evaluate on the ResNet-18~\cite{resnet18} in our main experiments.}
To study the transferability of the proposed DH scheme, we consider models with diverse architectures, including ResNet-50~\cite{resnet18}, VGG-19~\cite{VGG}, DenseNet-121~\cite{densenet}, and ViT~\cite{VIT}.

{\subsubsection{DH model training}
Our training exclusively utilizes the ImageNet subset comprising 100 classes for the DH model.
As detailed in Section~\ref{image_gen}, for each class, we generate 100 semantic images using paired text prompts and canny edge maps.
Our training configuration is as follows: 5k iterations, the Adam optimizer~\cite{adam} with hyper-parameters set at $\beta_{1}=0.5$, $\beta_{2}=0.999$, and $\epsilon=10^{-6}$; a consistent learning rate of $1 \times 10^{-4.5}$; and a mini-batch size of 24. 
To ensure stable model training, we assign weights of 1 to $\omega_1$ and $\omega_2$, and a weight of 0.0001 to $\omega_3$ across all experiments.
Subsequent to this, the pre-trained DH model is used to {generate unlearnable examples} across the three datasets: CIFAR-10, CIFAR-100, and the ImageNet subset. 
{Under the premise of ensuring invisibility, we further adopted a strict boundary to clip the scale of perturbations within $8/255$.}
The unlearnable examples generation follows two settings: \textit{class-wise setting} and \textit{sample-wise setting}.  
In the class-wise setting, we hide a consistent semantic image in the clean images of each class; whereas in the sample-wise setting, the hidden semantic images in each class are sampled from the generative model with the same text prompt and canny edge map, so they share the same global shape but differ in their local textures.
}

{\subsubsection{Classifier training}
To evaluate the effectiveness of the generated unlearnable examples, we employ a standard classification problem. For both CIFAR-10 and CIFAR-100 datasets, we follow the official train-test division. Regarding the ImageNet subset, we allocate 20\% of images from the first 100 classes of the official ImageNet training set for training purposes, using all related images in the official validation set for testing. We perform 40,000 iterations on the CIFAR-10 and CIFAR-100 datasets, and 8,000 iterations on the ImageNet subset. 
}

\subsubsection{Compared methods} We compare the proposed deep hiding scheme with six state-of-the-art unlearnable examples methods, including four surrogate-dependent methods, EM~\cite{EM}, REM~\cite{REM}, TAP~\cite{TAP}, and {EntF~\cite{EntF}}, and three surrogate-free methods, AR~\cite{AR}, LSP~\cite{LSP}, and OPS~\cite{OPS}. {The perturbations from EM, REM, TAP, and EntF are constrained by an $L_{\infty}$ norm of 8; AR and LSP use $L_{2} = 1$; OPS is $L_{0} = 1.0$. Additionally, to assess whether other hiding methods used in backdoor attacks can generate unlearnable examples, we tested the performance of methods including Poison Ink~\cite{zhang2022poison} and AdvINN~\cite{chen2023imperceptible}. AdvINN is implemented using the proposed Classifier Guided Target Image (CGT) strategy to execute attacks, according to their public codes. Note that we re-implemented all methods based on the publicly available codes.}
\subsubsection{Robustness evaluation} 
{To evaluate the robustness of our generated unlearnable examples, we train ResNet-18 with them across a variety of data augmentations and preprocessing methods, as suggested in~\cite{ISS}.
For augmentation, we employ vanilla (random crop and resize), cutout~\cite{cutout}, cutmix~\cite{cutmix}, and mixup~\cite{mixup}.
Additionally, we utilize seven data preprocessing techniques: Mean Filter (MeanF), Median Filter (MedianF), Bit-Depth Reduction (BDR), Grayscale transformation (Gray), Gaussian Noise (GaussN), Gaussian Filter (GaussF), and JPEG compression.
{Additionally, we also implement adversarial training (AT)~\cite{madry2017towards} with the constraint on both $L_2$  and $L_{\infty}$  norms. }
In alignment with the settings in REM~\cite{REM}, the specific details are as follows.

\begin{itemize}
\item {Basic Augmentation (Vanilla).} For CIFAR-10 and CIFAR-100 datasets, our data augmentation comprises random flipping, padding by 4 pixels on each side, followed by random cropping to a size of $32 \times 32$. Each image's pixels are then rescaled to the range $\displaystyle [-0.5, 0.5]$. In the case of the ImageNet subset, we augment the data using random cropping, resizing the images to a $224 \times 224$ dimension, implementing random flipping, and then rescaling every pixel to the interval $\displaystyle [-0.5, 0.5]$. 

\item {Cutout.} We adjust the sizes of the squared cutout box to match the image sizes of the datasets: $16$ for CIFAR-10 and CIFAR-100, and $112$ for the ImageNet subset. The cutout box is randomly placed within each image and maintains a consistent size across all images.

\item {Cutmix.} For a given sample, we first generate a square bounding box centered at a randomly chosen position with a randomly selected size (ranging from 0 to 1). The content within this bounding box is then replaced with content cropped from another randomly chosen image from the same mini-batch.

\item {Mixup.} We randomly select another image and blend it with the current sample using a randomly chosen weight (ranging from 0 to 1). We retain the original label for the current sample during loss function computation.

\item {Filters.} We use a kernel size of 3 for median, mean, and Gaussian smoothing (with a standard deviation of 0.1).

\item {Bit-Depth Reduction (BDR).} We implement 2 bits to perform BDR transformation.

\item {Grayscale.} We first calculate the weighted sum of the three channels and then replicate it across all three channels.

\item {Gaussian Noise.} We generate noise for each sample with a distribution of $\mathcal{N}\left(0, 0.1\right)$.

\item {JPEG Compression.} JPEG compression qualities set at 10\% and 50\%. 

\item {Adversarial Training (AT)~\cite{madry2017towards}.} 
{The $L_{\infty}$ and $L_2$ norm-bounded perturbations of scale $8/255$ and 1, respectively. }PGD-10 is employed with a step size of $2/255$, training the model on CIFAR-10 for 100 epochs.

\end{itemize}

{In addition to these common countermeasures, we also evaluate the robustness of our method against more contemporary countermeasures using their public codes, including Orthogonal Projection (OP)~\cite{sandoval2024can}, UEraser~\cite{qin2023learning}, and AVATAR~\cite{AVATAR}. Furthermore, we assess the robustness of our method in an unsupervised contrastive learning scheme, \ie, SimCLR~\cite{he2022indiscriminate}. }

\begin{table*}[t]
\centering
%\tabcolsep=0.25cm
%\scriptsize
\renewcommand\arraystretch{1.15}
\caption{\mrh{Test accuracy (\%) of models trained on unlearnable examples from CIFAR-10, CIFAR-100, and ImageNet subset against data augmentations, data preprocessing, and adversarial training. {Numbers} in {Bold} and {Underline} numbers indicate the best and second-best results, respectively.}}

\label{Robust}
\resizebox{1.01\linewidth}{!}{
%\begin{tabular}{c|c|cccc|cccccccc|c|c}
\begin{tabular}{c|c|c|p{0.6cm}p{0.6cm}p{0.6cm}p{0.6cm}|p{0.6cm}p{0.7cm}p{0.65cm}p{0.65cm}p{0.65cm}p{0.65cm}p{0.65cm}p{0.8cm}|p{0.8cm}p{0.65cm}|cc}
\hline
&{Norm}& Method & Vanilla & Cutout  & Cutmix  & Mixup   & MeanF& MedianF & BDR & Gray & GaussN & GaussF & JPEG10 &JPEG50 & AT($L_{\infty}$) & AT($\ell_2$) & Mean & Max\\
\hline
\hline
\multirow{9}{*}{\rotatebox{90}{CIFAR-10}} 
&-&Clean & 94.59 & 95.00   & 94.77 & 94.96  & 49.70   & 86.64   & 89.07   & 92.80  & 88.71   & 94.54   & 85.22 & 90.89& 84.19 & 83.54 & 87.47 & 95.00\\
\cline{2-19}
% \hline
&\multirow{6}{*}{$L_{\infty}$}
&EM~\cite{EM}& \underline{10.10}   & \textbf{10.00}   & 15.39   & 16.82   & {10.63}   & 24.27   & 35.90   & 69.29   & 32.96   & \textbf{10.01}   & 84.80  & 87.82& 84.28 & 83.33 & 41.11 & 87.82\\

&&REM~\cite{REM}  & 29.00   & 29.42   & 26.13   & 28.37   & 19.07   & 32.80   & 39.93   & 69.83   & 39.97   & 28.67   & 84.15&77.65   & 85.93 & 84.87 & 48.27 & 85.93\\

&&TAP~\cite{TAP} & 25.90   & 32.69   & 26.77   & 40.46   & 31.68   & 65.12   & 80.25   & 26.36   & 88.66   & 26.09   & 84.77  &90.31 & 83.57 & 82.74&  56.10 & 90.31\\

&&{EntF~\cite{EntF}} & 91.50  & 91.30 &90.93& 92.52 & 17.85 & 70.28&91.46 &80.33& 90.31& 79.79  & \textbf{74.36} &83.56 &\underline{75.86} & \underline{74.05} &  78.86 & 92.52\\

&&DH(S) & 12.93 & 14.03 & 13.80 & 15.89 & 16.06 & 25.56 & \underline{28.28} & 25.65 & 65.11 & \underline{12.30} & 84.93 & 89.18 & 83.44 & 83.27& \underline{37.47} & 89.18\\

&&DH(C) & \textbf{10.08} & \textbf{10.00} & \textbf{10.81} & \textbf{10.05} & \textbf{10.20} & \textbf{15.58} & \textbf{17.31} & \textbf{10.32} & \underline{30.23} & \textbf{10.01} & 83.98 & 78.96 & 82.21 & 82.34& \textbf{33.01} & \textbf{83.98}\\
\cline{2-19}

&$L_0$&OPS~\cite{OPS} & 16.53   & 89.73   & 83.91   & 34.88   & 17.31   & 86.86   & 43.04   & \underline{16.65} & 36.72   & 15.10 & \underline{82.79} &\textbf{57.00}  & \textbf{9.42} &\textbf{65.60} & 46.82 & 89.73\\
\cline{2-19}

&\multirow{2}{*}{$L_2$}&LSP~\cite{LSP} & 19.07   & 19.87   & 20.89   & 26.99   & 28.85   & 29.85   & 66.19   & 82.47   & \textbf{19.25}   & 16.19   & 83.01 &\underline{57.87}   & 84.59 & 83.96 & 45.65 & \underline{84.59}\\

&&AR~\cite{AR} & 13.31   & \underline{11.35}   & \underline{12.21}   & \underline{13.30}   & \underline{12.38}   & \underline{17.04}   & 37.42   & 34.81   & 42.29   & 12.56   & 85.08&  89.63& 83.17
& 84.29& 39.20 & 89.63
%\ud{58.23} %37.42
\\
\hline
\hline

\multirow{9}{*}{\rotatebox{90}{CIFAR-100}} 
&-& Clean & 75.82 & 74.45 & 76.32 & 77.07  & 14.72&50.72 & 63.51 & 70.04 & 62.41& 75.86 &57.35& 68.59 &58.25 & 57.40& 63.04 & 77.07\\
\cline{2-19}

&\multirow{6}{*}{$L_{\infty}$} &EM~\cite{EM} & 2.84 & 12.05 & 7.67& 12.86 & 13.52 & 43.61& 62.12& 62.37 & 62.01& 73.47& 57.29& 67.50&57.89 & 56.63& 42.27 & 73.47\\

& &REM~\cite{REM} & 7.13 & 10.32 & 11.25 & 8.65& 5.90& 12.31& 19.95& 48.48  & 26.27& 7.32 & 57.15& 65.10& 58.90 & 58.75& 28.39 & 65.10\\

& &TAP~\cite{TAP} & 14.00  & 16.55 & 15.99 & 22.56 & 5.86& 31.95& 55.12& \underline{8.90} & 61.40 & 13.95& 56.56& 66.67& 56.53 & 55.53& 34.40 & 66.67\\

& &{EntF~\cite{EntF}} & 72.55  & 69.65 & 70.68 & 73.81 & 8.67 & 36.87& 55.22& 67.00 &58.54 &73.10 & \textbf{51.42} & 63.69 & \underline{52.44} & \underline{50.66} & 57.46 & 73.81\\

&&DH(S) & 7.81 & 4.80 & 10.15 & 10.27 & 10.14 & 22.23 & 38.13 & 15.50 & 52.43 & 7.97 & 56.01& 65.72 & 56.48 & 56.38& 29.57 & 65.72\\

&&DH(C) & \textbf{1.22} & \textbf{1.01} & \textbf{1.22} & \textbf{1.09} & \textbf{1.51} & \textbf{2.72} & \textbf{12.08} & \textbf{0.96} & \underline{19.86} & \textbf{1.01} & \underline{55.07} & 53.83 & 56.91 & 56.34& \textbf{18.91} & \textbf{56.91}\\
\cline{2-19}

&$L_0$&OPS~\cite{OPS} & 11.69 & 71.36 & 64.25 & 12.59 & \underline{3.18} & 49.74& \underline{19.31} & 18.70& \textbf{17.30} & 11.79& 56.72& \underline{48.71} & \textbf{10.22} & \textbf{48.24}& 31.70 & {64.25}\\
\cline{2-19}

&\multirow{2}{*}{$L_2$}&LSP~\cite{LSP} & 2.68& 2.55& 2.69& 4.39& 7.15& 6.76 & 28.23& 42.77& 22.42& 2.19 & 55.23 &\textbf{33.60}& 57.45 & 56.32& \underline{23.17} & \underline{57.45}\\

&&AR~\cite{AR} & \underline{1.50} & \underline{1.47}& \underline{1.56} & {1.37}& 5.35& \underline{3.89} & 28.28& 19.68& 59.34& \underline{1.57} & 56.99 &65.72& 58.33 & 56.60& 25.83 & 65.72\\
\hline
\hline
\multirow{7}{*}{\rotatebox{90}{ImageNet subset}} &-& Clean & 63.93& 64.02 & 55.10 &  64.55 &19.92 & 36.08 &56.63 &68.35  &50.62 &65.40 & 56.83 & 69.36 &48.24 & 62.32& 55.81 & 69.36\\
\cline{2-19}
&\multirow{5}{*}{$L_{\infty}$}&EM~\cite{EM}& 28.99 & 18.78   & 17.61  & 36.55  &  7.46 & 32.60 & 53.43  & 17.93   & 44.63 & 26.04 &  53.41 & 56.96 & {43.56}& 42.26 & 34.30 & 56.96\\
&&REM~\cite{REM} &  14.78 & 14.10   & 11.73   & 19.88  & 15.32   & 14.12 & 16.48  &  44.74 & \textbf{15.96} & 15.34 &50.50 &\underline{17.14}&47.52 & \textbf{21.10}& 22.76  & 50.50\\
&&TAP~\cite{TAP} &  7.96 & 15.02& 15.18   & 23.08  & 10.44  &15.02  & 47.97  & 22.93  & 46.84 & 12.80 & 53.40 & 37.98& 44.18 & 41.56& 28.17 & 53.40\\
&&DH(S) & \underline{2.70} & \underline{2.98} & \textbf{1.86} & \underline{3.74} & \underline{3.44} & 18.18 & \underline{10.70} & \underline{2.22} & \underline{19.04} & \underline{2.58} & \underline{27.84} & 18.02 & 44.24
& 30.76& \underline{13.45} & \underline{44.24}
%11.31
\\
&&DH(C) & \textbf{2.42} & \textbf{2.30} & \underline{3.28} & \textbf{1.94}  & \textbf{2.38}  & \textbf{3.78}  & \textbf{8.44}  & \textbf{1.38} & {27.48}   & \textbf{1.84}  & \textbf{25.12}  & \textbf{10.24} & \textbf{42.14}
& \underline{26.66}& \textbf{11.39} & \textbf{42.14}\\
\cline{2-19}
&\multirow{1}{*}{$L_2$}&LSP~\cite{LSP}  & 18.18 & 9.52   & 34.16   & 9.76  & 4.14 & \underline{5.20}   & 43.38 & 52.66   & 34.28   & 17.92    & 51.80 & 49.06 & \underline{42.26} & 41.08& 29.53 & 52.66\\
\hline
\end{tabular}
}
\end{table*}

\subsection{{Invisibility of the proposed method}}
\label{sec:overall_performance}
%{Effectiveness of UEs.} 
{In this part, invisibility is first assessed using both subjective and objective metrics. For subjective evaluation, we visualize the generated unlearnable examples with their perturbations on the ImageNet subset using our hiding scheme, as shown in Figure \ref{Fig:perturbation}.}
{
Regarding invisibility, our generated unlearnable examples are more imperceptible compared to those from other methods. Additionally, our generated perturbations consistently display high-level features, such as shape, that align with the hidden semantic image.
For objective evaluation, we assess the image quality using PSNR between the generated unlearnable examples and their corresponding clean images to analyze the perturbation scale on the ImageNet subset. The results of PSNR  are as follows: DH(C) at 36.25, EM at 34.17, REM at 33.96, TAP at 35.86, and LSP at 23.59. We observe that with a perturbation scale of $\epsilon = 8/255$ under the $L_{\infty}$ norm, the proposed DH method achieves superior image quality. This improvement is due to DH is an adaptive hiding process through INN. In contrast, existing methods that generate unlearnable examples based on the image domain typically restrict the scale of perturbations by the $L_{p}$ norm, resulting in uniformly distributed perturbations across the entire image.}
{These findings underscore the remarkable invisibility of our proposed DH method.}}

\subsection{Effectiveness of the proposed method}
\label{sec:overall_performance}
%{Effectiveness of UEs.} 
Under the strict bound and better invisibility, we evaluate the efficacy of the proposed method by training ResNet18 with unlearnable examples and testing on clean images. 
In Table~\ref{Robust}, we present detailed test accuracy results across three datasets: CIFAR-10, CIFAR-100, and the ImageNet subset. Notably, both our class-wise and sample-wise settings consistently achieve the best performance on all three datasets.
{Specifically, the results of the vanilla training show that our class-wise setting degrades the test accuracy to 10.08\%, 1.22\%, and 2.42\% for three datasets respectively,  which are nearly random-guessing. }
It indicates that the models can not learn any useful semantic information for the classification task once we hide the semantic images into clean images. 
In contrast, the other unlearnable examples techniques fail to maintain consistent performance across datasets. 
For instance, both EM and LSP result in much higher test accuracies on ImageNet.
{Even though we use a sample-wise setting to reduce the exposure risk of the hidden image, it still achieves promising performance across datasets, especially on ImageNet.} 
We hypothesize that our unlearnable examples carry abundant information due to their semantic image nature, making them more generally effective in various scenarios.

\begin{figure*}
\begin{center}
\includegraphics[width=0.95\textwidth]{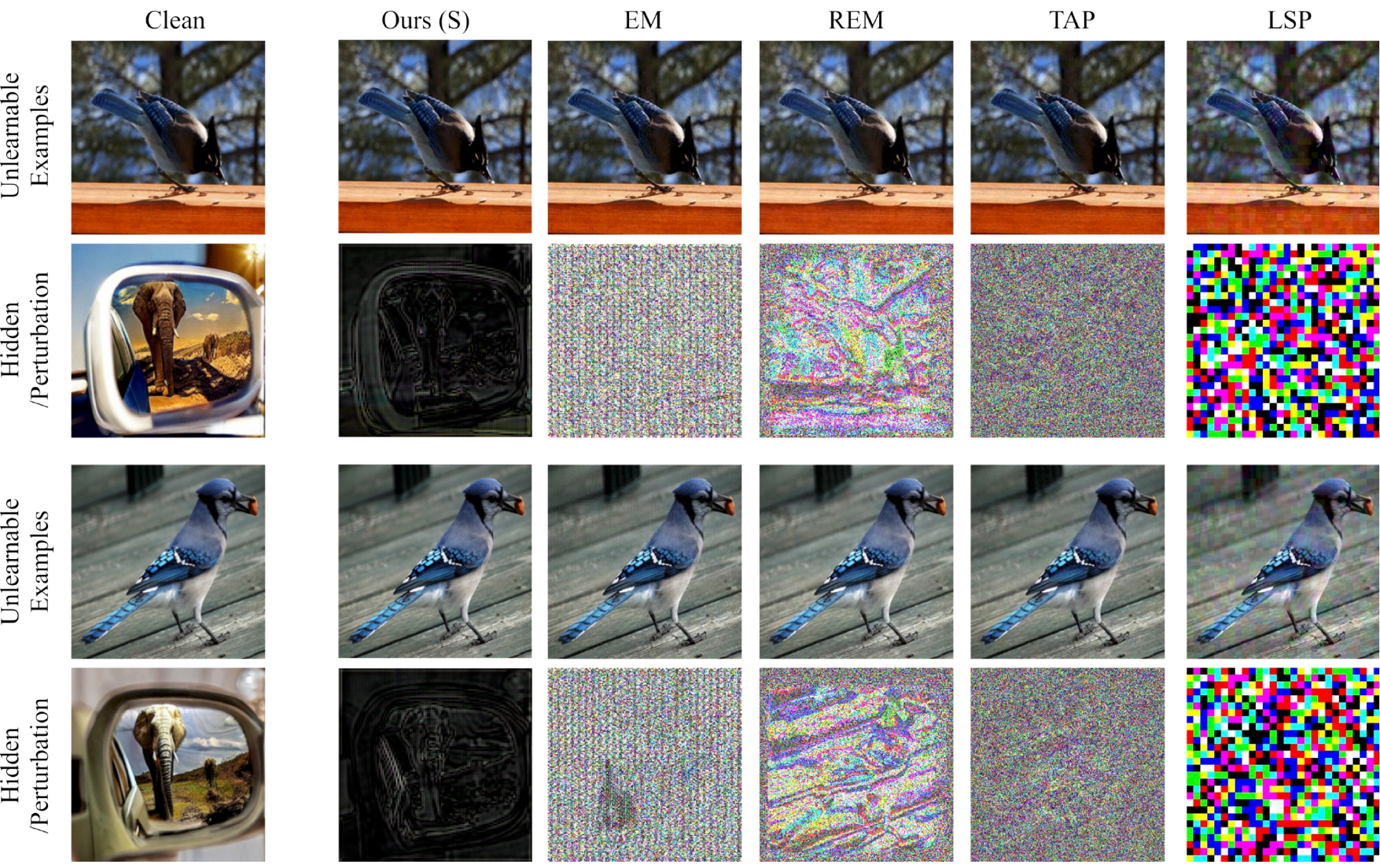}
\vspace{-1mm}
\caption{
%Visualization of perturbation maps.
{The visualization of the unlearnable examples generated by different methods is shown in columns 2-5, where the second and fourth rows correspond to the perturbation maps.
Perturbations are absoluted and normalized to [0,1] for a better view.}}
\label{Fig:perturbation}
\end{center}
\end{figure*}

\iffalse
\begin{table}
\centering
\renewcommand\arraystretch{1.15}
\caption{Test accuracy (\%) of CIFAR-10 on the models trained by the clean data mixed with different percentages of unlearnable examples.}
\label{Tab:4}
\resizebox{\linewidth}{!}{
\begin{tabular}{c|p{0.5cm}p{0.5cm}p{0.5cm}p{0.5cm}p{0.5cm}p{0.5cm}p{0.5cm}p{0.5cm}}
\hline
Method   & \multicolumn{1}{r}{20\%}  & \multicolumn{1}{r}{40\%}  & \multicolumn{1}{r}{60\%}  & \multicolumn{1}{r}{80\%} & \multicolumn{1}{r}{90\%}& \multicolumn{1}{r}{95\%}& \multicolumn{1}{r}{99\%}& \multicolumn{1}{r}{100\%}  \\ \hline
EM~\cite{EM}& 94.30 & 93.09 & 91.42 & 87.29 & 82.53 & 75.88 & 58.25 & 10.1 \\
REM~\cite{REM} & 93.83 & 92.69 & 91.12 & 86.92 & 81.67 & 75 & 52.39 & 30.4  \\
TAP~\cite{TAP} & 93.82 & 92.78 & 91.96 & 88.49 & 86.8 & 82.52 & 72.88 & 25.93 \\
{EntF~\cite{EntF}}&  93.40 & 91.71 &  91.25 &91.07 & ...&...&...&...\\
LSP~\cite{LSP} & {93.50} & {92.47} & 90.21 & 84.81 & 82.52 & 74.37 & 51.7 & 16.99 \\
OPS~\cite{OPS} & 93.64 & 92.63 & {90.05} & {84.42} & 83.25 & 77.5 & 59.9 & 11.88 \\
AR~\cite{AR} & 94.07 & 92.66 & 90.34 & 85.18 & 82.77 & 71.48 & 41.59 & 17.46\\
DH(S) & {93.53} & 92.67 & {89.99} & 84.47 & ...&...&...&...\\
DH(C) & 93.73 & {92.41} & 90.08 & {84.40} & 80.5 & 73.05 & 31.17 & 10\\ \hline
\end{tabular}}
\end{table}
\fi

\begin{table}[t]
\centering
%\tabcolsep=0.25cm
%\scriptsize
\renewcommand\arraystretch{1.15}
\caption{\mrh{Test accuracy (\%) of models trained on unlearnable examples from CIFAR-10 against more contemporary countermeasure. {Numbers} in {Bold} and {Underline} numbers indicate the best and second-best results, respectively.}}
% \vspace{5pt}
% Hyperparameters for different countermeasures can be found in Appendix ?.
\label{Robust2}
\resizebox{1.0\linewidth}{!}{
\begin{tabular}{c|c|ccc|c|cc}
% \begin{tabular}{c|c|c|p{0.6cm}p{0.6cm}p{0.6cm}p{0.8cm}|c}
\hline
{Norm}& Method & OP & UEraser & AVATAR & SimCLR  &  Mean & Max\\
%\hline
\hline
\multirow{9}{*} 
&Clean & 90.16 & 92.88  & 90.35 & 74.03 & 86.86  & 92.88\\
\cline{1-8}
% \hline
\multirow{6}{*}{$L_{\infty}$}
&EM~\cite{EM}& 65.17 & 45.52  & 90.95 & 71.66  & 68.33 & 90.95\\
&REM~\cite{REM} & 25.83 & 63.18 & 88.49 & \textbf{68.59}   & 61.52 & 88.49\\
&TAP~\cite{TAP} & 14.74  & 76.21 & 90.71& 72.01   & 63.42 & 90.71 \\
&{EntF~\cite{EntF}} & 88.43 & 90.73  & 90.67 & 71.39  & 85.31 & 90.73\\
&DH(S) & 14.44 & \underline{25.54}  & 88.69 & 72.70   & \underline{50.34} & 88.69\\
&DH(C) & \textbf{10.00} & \textbf{20.52}  & 87.38 & \underline{71.27}  & \textbf{47.29} & \textbf{87.38}\\
\cline{1-8}
$L_0$&OPS~\cite{OPS} & 87.94 & 72.52   & \textbf{66.16} & 71.42 & 74.51 & \underline{87.94}\\
\cline{1-8}
\multirow{2}{*}{$L_2$}
&LSP~\cite{LSP} & 87.99 & 85.07  & \underline{85.69} & 72.27  & 82.76 & 87.99\\
&AR~\cite{AR} & \underline{13.03} & 93.16  & 91.57 & 73.33   & 67.77 & 93.16\\
\hline
\end{tabular}
}
\end{table}

\vspace{0.4cm}
\subsection{Robustness of the proposed method}
%{Robustness of UEs.} 
To evaluate the robustness of our generated unlearnable examples, we adopt a variety of common countermeasures, including four data augmentation, seven data preprocessing techniques, {and AT with constraints on both $L_2$  and $L_{\infty}$  norms.}
As shown in Table~\ref{Robust}, the experimental results demonstrate that the proposed method consistently outperforms the other techniques, 
{exhibiting robust performance against most countermeasures.}
On CIFAR-10 dataset, our method reduces the test accuracy to 10\%$\sim$17.31\% across a broad range of countermeasures. Despite that, DH shows less robustness against JPEG compression and AT, similar to the majority of other methods except OPS and EntF. OPS modifies only one pixel, which might be less affected by JPEG compression and could remain unnoticed in AT scenarios. However, it inherently has trouble with the cropping and filtering operation, leading to diminished results in scenarios like cutout, cutmix, and median filtering. As for EntF, although it demonstrates relative robustness to AT, it is less effective against other countermeasures.
This highlights a trade-off where some methods may excel in specific conditions but falter in others. As for mean test accuracy, DH outperforms others with the lowest value noted at 6.19\%. This represents a significant improvement over the next best method, AR, and suggests that DH is generally more effective at generating robust unlearnable examples against various countermeasures. 
\mrh{Additionally, for maximum test accuracy, which represents the highest performance among all countermeasures, DH still demonstrates superior performance, indicating its potential efficacy in real-world scenarios.
}
The experimental results on CIFAR-100 dataset are similar to those on CIFAR-10. Regarding the mean values, our method still outperforms LSP by about 4\% and exceeds EntF by nearly 40\% on CIFAR-100. 
\mrh{Furthermore, when considering the maximum values, our method consistently achieves the best performance.} This highlights the superiority of our proposed method.
{For ImageNet subset, DH shows more significant improvements compared to other methods and demonstrates general robustness. 
We speculate that the robustness of ImageNet stems from its high resolutions, which are capable of carrying a richer array of semantic image information.
Additionally, since the hiding model is trained on the ImageNet dataset, the effectiveness of hiding is also improved.}
\mrh{Overall, by observing the mean and maximum robustness assessment values, our method has achieved the best performance against common countermeasures.}
{Specifically, we obtain mean test accuracy of 33.01\%, 18.91\%, and 11.39\% on CIFAR-10, CIFAR-100, and ImageNet subset, respectively, 
compared to the best performances of other methods at 39.20\%, 23.47\%, and 22.76\%.} 
\mrh{Additionally, our method attained maximum test accuracies of 83.98\%, 56.91\%, and 42.14\% on CIFAR-10, CIFAR-100, and ImageNet subset, respectively, 
compared to the best performances of other methods at 84.59\%, 57.45\%, and 44.24\%.}
% \mrh{To be more relevant to real-world scenarios, we also report the highest test accuracy among all countermeasures on diverse datasets. It shows that DH shows the best performance.}
These results represent that our deep hiding scheme obtains a better robust generalization of unlearnable examples with high-level semantic features.

{In addition to evaluating the performance of our proposed method against common countermeasures, we have also tested it against some more contemporary countermeasures. The results, as shown in Table~\ref{Robust2}, indicate that our method effectively thwarts OP~\cite{sandoval2024can} and UEraser~\cite{qin2023learning}, maintaining exceptionally low test accuracies. Additionally, its performance is comparable to other methods against AVATAR~\cite{AVATAR}. We further assess the robustness of our method in an unsupervised contrastive learning setting, such as SimCLR~\cite{he2022indiscriminate}, and find our results comparable to those of other methods. Additionally, it demonstrates that our method achieves the second-best performance compared to others. Notably, our method demonstrates strong performance against OP with test accuracies of 14.44\% and 10.00\%. 
This is attributed to our adaptive hiding manner, which is tailored to the unique features of each clean image. Such variability in perturbations enables our method to effectively resist OP. Overall, the mean and maximum results indicate that our method has better robustness performance compared to other methods.}

\begin{table*}
\centering
\renewcommand\arraystretch{1.15}
\caption{{Test accuracy (\%) of CIFAR-10 and CIFAR-100 on five architectures, including  ResNet-18 (R18), ResNet-50 (R50), VGG-19 (V19), DenseNet-121 (D121), and Vision Transformer (ViT).{Numbers} in {Bold} and {Underline} numbers indicate the best and second-best results, respectively.}}
\label{Tab:3}
\resizebox{0.7\linewidth}{!}{
\begin{tabular}{c|c|p{0.6cm}p{0.6cm}p{0.6cm}p{0.6cm}p{0.6cm}|p{0.6cm}p{0.6cm}p{0.6cm}p{0.6cm}p{0.6cm}}
\hline
\multirow{2}{*}{Norm}& Dataset & \multicolumn{5}{c|}{CIFAR-10}  & \multicolumn{5}{c}{CIFAR-100} \\ 
\cline{2-12}

\multirow{8}{*}{$L_{\infty}$} &Model & R18  & R50  & V19  & D121 & ViT & R18  & R50  & V19  & D121 & ViT \\ \hline
& EM~\cite{EM} & \textbf{10.10}  & \underline{10.00} & \textbf{10.82}  &  \underline{12.56} & \underline{11.88}  &2.84 & 3.88 & 9.23 & 64.87 & 7.65 \\
& REM~\cite{REM}& 30.40 & 25.10 & 24.54  & 30.28  &  32.36 &7.13 & 7.45 & 5.26 & 12.47 &   6.91 \\
& TAP~\cite{TAP} & 25.93 & 25.48 & 30.36  & 78.59  & 70.96 &14.00 & 14.25 & 33.18 &  52.64&  14.49 \\
& {EntF~\cite{EntF}}&  91.50 & 91.83 & 88.17 & 83.30  & 69.23 &72.55 & 73.19 & 65.68 & 60.85 & 49.43 \\
& DH(S)  & 12.93 & 10.64 & 12.69 & 20.52 & 65.70 &  7.81 & 11.51 & 14.55 & 14.36 & 12.48  \\
& DH(C)  & \underline{10.18} & \textbf{9.98} & 12.33 & \textbf{10.00} & \textbf{9.99} & \textbf{1.25} & \textbf{1.16} & \textbf{0.93} & \textbf{1.11} & \textbf{5.54}  \\
\hline

\multirow{1}{*}{$L_0$}& OPS~\cite{OPS}&  17.46 &  16.73& 19.12  & 18.40  &  28.09 & 11.69& 10.90 & 5.67 & \underline{10.38} &  17.23 \\
\hline

\multirow{2}{*}{$L_2$}& LSP~\cite{LSP}&  16.99 & 14.55 & \underline{11.53}  &  24.83 & 23.78 & 2.68& 4.06 &  \underline{2.84} &  27.05  &  9.40  \\
&AR~\cite{AR} &  11.88 & 15.83 & 13.21  & 22.28  & 19.84 &  \underline{1.50} &  \underline{2.13} & 3.48 & 19.55 &  \underline{5.69} \\
\hline
\end{tabular}}
\end{table*}

\begin{table*}[t]
\centering
\renewcommand\arraystretch{1.15}
\caption{
{Test accuracy (\%) of model train on unlearnable examples from CIFAR-10 with five architectures, including ResNet-18 (R18), ResNet-50 (R50), VGG-19 (V19), and DenseNet-121 (D121), and Vision Transformer (ViT), against data augmentations, data preprocessing, and adversarial training.}}
\label{across}
%\resizebox{\linewidth}{!}{
\begin{tabular}{c|c|p{0.6cm}p{0.6cm}p{0.6cm}p{0.6cm}|p{0.6cm}p{0.7cm}p{0.6cm}p{0.6cm}p{0.6cm}p{0.6cm}p{0.6cm}p{0.8cm}|p{0.7cm}p{0.7cm}|c}
\hline
 & Model & Vanilla & Cutout  & Cutmix  & Mixup& MeanF& MedianF & BDR & Gray & GaussN & GaussF & JPEG10 &JPEG50 & AT($L_{\infty}$) & AT($\ell_{2}$) & Mean\\
\hline
\multirow{5}{*}{\rotatebox{90}{Ours(S)}} 
& R18  & 12.93 & 14.03 & 13.80 & 15.89 & 16.06 & 25.56 & 28.28 & 25.65 & 65.11 & 12.30 & 84.93 & 89.18 & 83.44 & 83.27& 37.47  \\
&R50 & 10.64 & 13.26 & 13.81 & 22.40 & 17.36 & 57.97 & 72.39 & 22.92 & 82.43 & 11.06 & 84.69 & 89.38 & 85.26 & 84.59 & 47.72 \\
&V19& 12.69 & 12.31 & 12.58 & 18.60 & 29.63 & 54.55 & 64.13 & 25.34 & 77.49 & 13.04 & 83.38 & 88.04 & 77.87 & 79.13 & 46.34 \\
&D121& 20.52 & 27.57 & 28.29 & 25.56 & 73.59 & 81.89 & 75.34 & 55.51 & 72.16 & 21.91 & 82.13 & 85.03 & 75.66 & 76.78 & 57.29 \\
&ViT & 65.70 & 69.02 & 59.43 & 65.30 & 65.87 & 69.27 & 67.42 & 44.95 & 66.08 & 75.97 & 67.16 & 73.16 & 43.57 & 50.49 & 63.86 \\
\hline\hline

\multirow{5}{*}{\rotatebox{90}{Ours(C)}} 
& R18 &  10.08 & 10.00 & 10.81 & 10.05 & 10.20 & 15.58 & 17.31 & 10.32 & 30.23 & 10.01 & 83.98 & 78.96 & 82.21 & 82.34& 33.01  \\
&R50 & 9.89 & 10.02 & 12.93 & 10.23 & 10.40 & 17.06 & 17.37 & 10.02 & 24.29 & 9.96 & 84.70 & 76.42 & 83.40 & 83.37 & 32.86\\
&V19& 12.33 & 9.61 & 10.48 & 10.39 & 10.75 & 15.71 & 13.68 & 11.14 & 27.99 & 10.72 & 83.91 & 78.93 & 79.67 & 79.53 & 32.48\\
&D121& 10.00 & 14.06 & 14.78 & 12.59 & 55.75 & 59.64 & 25.59 & 15.93 & 30.66 & 10.01 & 81.83 & 74.82 & 76.19 & 77.31 & 39.94 \\
&ViT & 9.99  & 10.30 & 10.35 & 42.85 & 64.08 & 62.49 & 43.88 & 15.32 & 23.36 & 10.07 & 66.55 & 66.84 & 44.44 & 49.70 & 37.20\\
\hline
\end{tabular}
%}
\end{table*}

\subsection{Performance on different architectures and unlearnable percentages}

\subsubsection{Different model architectures} In real-world scenarios, the protector may not know the details of the target model. 
{In such cases, it's critical for unlearnable examples to be transferable.
Hence, we evaluate the effectiveness of the proposed method across various deep learning architectures on CIFAR-10 and CIFAR-100 datasets.}
As shown in Table \ref{Tab:3}, 
{our approach  in the class-wise setting consistently performs well across all five models.}
To further validate the effectiveness of our proposed methods, we conducted more comprehensive cross-validation. We verify the robustness of our generated unlearnable examples against various countermeasures under different architectures, and the results are shown in Table \ref{across}. It supports our claim that the proposed DH maintains its efficacy against different countermeasures across architectures.

\begin{table}[t]
\centering
\renewcommand\arraystretch{1.20}
\caption{{Test accuracy (\%) of CIFAR-10 on the models trained by the clean data mixed with different percentages of unlearnable examples. {Numbers} in {Bold} and {Underline} numbers indicate the best and second-best results, respectively.}}
\label{Tab:4}
\resizebox{0.78\linewidth}{!}{
\begin{tabular}{c|c|p{0.5cm}p{0.5cm}p{0.5cm}p{0.5cm}}
\hline
Norm & Method   & \multicolumn{1}{r}{20\%}  & \multicolumn{1}{r}{40\%}  & \multicolumn{1}{r}{60\%}  & \multicolumn{1}{r}{80\%}   \\ \hline
\multirow{6}{*}{$L_{\infty}$} & EM~\cite{EM}& 94.30 & 93.09 & 91.42 & 87.29  \\
& REM~\cite{REM} & 93.83 & 92.69 & 91.12& 86.92   \\
& TAP~\cite{TAP} & 93.82 & 92.78 & 91.96 & 88.49  \\
& {EntF~\cite{EntF}}&  \underline{93.40} & \textbf{91.71} &  91.25 &91.07 \\
& DH(S) & 93.79 & 92.64 & 91.14 & 86.29 \\
& DH(C) &\textbf{93.33} &\underline{92.36} &\underline{90.09} & \textbf{83.99} \\ 
\hline

\multirow{1}{*}{$L_0$} & OPS~\cite{OPS} & 93.64 & 92.63 & \textbf{90.05} & \underline{84.42} \\
\hline
\multirow{2}{*}{$L_2$} & LSP~\cite{LSP} & {93.50} & {92.47} & 90.21 & 84.81  \\
& AR~\cite{AR} & 94.07 & 92.66 & 90.34 & 85.18 \\
\hline
\end{tabular}}
% \vspace{-0.4cm}
\end{table}

\begin{table}[t]
\centering
\renewcommand\arraystretch{1.15}
\caption{{Test accuracy (\%) of CIFAR-10 on the different models trained by 
the clean data mixed with different percentages of unlearnable examples.}}
\label{across_p}
\resizebox{0.75\linewidth}{!}{
\begin{tabular}{c|c|p{0.5cm}p{0.5cm}p{0.5cm}p{0.5cm}}
\hline
Setting   & Model& {20\%}  & {40\%}  & {60\%}  & {80\%}  \\ 
\hline
\multirow{5}{*}{\rotatebox{90}{DH(S)}}&R18&93.79&92.64&91.14&86.27\\
&R50&94.38&94.00&90.90&86.83\\
&V19&92.88&92.52&88.24&83.13\\
&D121&88.98&89.72&89.27&83.97\\
&ViT&76.41&76.33&77.24&76.08\\
\hline\hline

\multirow{5}{*}{\rotatebox{90}{DH(C)}}&R18&93.33&92.36&90.09&83.99\\
&R50&93.38&92.56&90.42&82.32\\
&V19&91.70&90.83&88.32&80.35\\
&D121&89.40&86.83&83.83&78.76\\
&ViT&76.05&75.56&75.86&69.46\\

\hline
\end{tabular}}
% \vspace{-0.4cm}
\end{table}

\subsubsection{Different unlearnable percentages} 
Consider a situation where it's not feasible to protect all the data. 
This scenario is realistic since a practitioner who gains access to unlearnable examples might also obtain additional clean data from other avenues. 
{Consequently, it's common practice to evaluate unlearnable examples' efficacy by training deep learning models with a random subset of unlearnable examples.
To this end, we evaluate the performance of our proposed approach by using varying mixtures of clean images and unlearnable examples, the results are shown in Table \ref{Tab:4}. 
}
Besides, we also conducted the transferability studies across architectures with limited unlearnable examples, and the results as shown in Table \ref{across_p}. The test accuracy decreases in a similar trend when we increase the percentage of the unlearnable examples.

\subsection{Ablation study}

\begin{table*}[t]
\centering
\renewcommand\arraystretch{1.15}
\caption{{Ablation studies on CIFAR-10 for designed Latent Feature Concentration module (LFC), and Semantic Images Generation module (SIG), including Text Prompts Clustering (TPC) and Stable Diffusion model and ControlNet (SD+C). {Numbers} in {Bold} indicate the best and second-best results, respectively.}}
\label{Tab:Ablation}
\resizebox{1.0\linewidth}{!}{
\begin{tabular}{p{0.15cm}|p{0.34cm}|p{0.3cm}p{0.55cm}|p{0.6cm}p{0.6cm}p{0.6cm}p{0.6cm}|p{0.6cm}p{0.7cm}p{0.65cm}p{0.65cm}p{0.65cm}p{0.65cm}p{0.65cm}p{0.8cm}|p{0.8cm}p{0.65cm}|c}
\hline
\multirow{2}{*}{ } & \multirow{2}{*}{LFC} &\multicolumn{2}{|c|}{SIG} & \multirow{2}{*}{Vanilla} & \multirow{2}{*}{Cutout} & \multirow{2}{*}{Cutmix} & \multirow{2}{*}{Mixup} & \multirow{2}{*}{MeanF} & \multirow{2}{*}{MedianF} & \multirow{2}{*}{BDR} & \multirow{2}{*}{Gray} & \multirow{2}{*}{GaussN} & \multirow{2}{*}{GaussF} & \multirow{2}{*}{JPEG10} & \multirow{2}{*}{JPEG50} & \multirow{2}{*}{ AT($L_{\infty}$)} & \multirow{2}{*}{AT($\ell_2$)} & \multirow{2}{*}{Mean} \\ \cline{3-4}

    & & \multicolumn{1}{|c}{TPC} & SD+C & & & & & & & & & & & & & & & \\

\hline
\multirow{5}{*}{\rotatebox{90}{DH(S)}}&$\times$&$\times$ & $\times$& 94.38 & 94.53 & 94.39 & 94.79 & 54.64 & 86.65 & 89.17 & 92.29 & 88.80 & 94.27 & 84.74 & 90.79 & 84.51 & 83.17 & 87.65
\\

&$\times$&$\checkmark$ & $\checkmark$& 10.58 & 22.75 & 16.51 & 24.69 & 36.95 & 79.03 & 75.43 & 19.48 & 87.81 & 16.78 & 84.34 & 90.83 & 84.15 & 82.57 & 52.28\\
&$\checkmark$&$\times$&$\checkmark$& 10.05 & 10.45 & 20.15 & 21.31 & 18.02 & 46.51 & 78.31 & 11.70 & 87.95 & 10.34 & 84.64 & 90.73 & 84.03 & 83.56 & 46.98\\
&$\checkmark$&$\times$&$\times$& 94.50 & 94.48 & 94.20 & 94.90 & 56.45 & 84.13 & 89.57 & 94.24 & 89.04 & 94.30 & 85.06 & 90.76 & 84.24 & 83.26 & 87.65\\

&$\checkmark$&$\checkmark$&$\checkmark$ & 12.93 & 14.03 & 13.80 & 15.89 & 16.06 & 25.56 & 28.28 & 25.65 & 65.11 & 12.30 & 84.93 & 89.18 & 83.44 & 83.27& 37.47\\
\hline\hline

\multirow{5}{*}{\rotatebox{90}{DH(C)}}
&$\times$&$\times$ & $\times$& 21.82 & 10.03 & 14.38 & 15.02 & 20.97 & 47.28 & 38.12 & 9.97 & 14.95 & 10.11 & 84.30 & 77.16 & 84.27 & 84.61 & 38.08
\\
&$\times$&$\checkmark$&$\checkmark$& 10.55 & 10.84 & 11.29 & 11.63 & 16.29 & 45.12 & 33.12 & 9.99 & 25.53 & 10.39 & 84.58 & 88.38 & 84.08 & 83.79 & 37.54\\
&$\checkmark$&$\times$&$\checkmark$& 10.13 & 13.47 & 11.15 & 14.79 & 24.11 & 19.87 & 36.42 & 10.49 & 22.47 & 9.89 & 84.67 & 90.25 & 84.02 & 83.64 & 36.81\\
&$\checkmark$&$\times$&$\times$& 9.99 & 10.01 & 10.41 & 9.55 & 26.59 & 19.74 & 24.61 & 10.02 & 13.10 & 9.99 & 84.83 & 86.24 & 84.06 & 84.34 & 34.53\\

&$\checkmark$&$\checkmark$&$\checkmark$ & 10.08 & 10.00 & 10.81 & 10.05 & 10.20 & 15.58 & 17.31 & 10.32 & 30.23 & 10.01 & 83.98 & 78.96 & 82.21 & 82.34& 33.01\\
 \hline
\end{tabular}}
\end{table*}

\subsubsection{Effectiveness of Latent Feature Concentration (LFC) module}
To understand the pivotal role of LFC in our approach, we conduct an ablation study focused on unlearnable performance.
The results are shown in Table~\ref{Tab:Ablation}. 
Whether in the class-wise or sample-wise settings, the introduction of LFC leads to a discernible reduction in testing accuracy across three datasets.
{Notably, the CIFAR-10 dataset under the median filter exhibits a sharp decline, from 79.03\% to 46.51\%.} 
These findings indicate that by focusing the latent feature of perturbations on intra-class characteristics, 
the unlearnability of the unlearnable examples is enhanced.

\begin{table*}[t]
\centering
\renewcommand\arraystretch{1.15}
\caption{{Ablation studies on CIFAR-10 for different settings on parameters of $\omega_1$, $\omega_2$, $\omega_3$. {Numbers} in {Bold} indicate the best and second-best results, respectively.}}
\label{Par_ablation}
\resizebox{1.0\linewidth}{!}{
\begin{tabular}{c|c|p{0.6cm}p{0.6cm}p{0.6cm}p{0.6cm}|p{0.6cm}p{0.7cm}p{0.65cm}p{0.65cm}p{0.65cm}p{0.65cm}p{0.65cm}p{0.8cm}|p{0.8cm}p{0.65cm}|c}
\hline
&{Setting}& Vanilla & Cutout  & Cutmix  & Mixup   & MeanF& MedianF & BDR & Gray & GaussN & GaussF & JPEG10 &JPEG50 & AT($L_{\infty}$) & AT($\ell_2$) & Mean\\
\hline

\multirow{4}{*}{$\omega_1$} &10&12.14&10.30&13.63&22.21&21.20&51.61&51.76&11.07&39.68&10.01&84.96&88.58
&84.49&83.81&41.81\\
&1(DH)& 10.08 & 10.00 & 10.81 & 10.05 & 10.20 & 15.58 & 17.31 & 10.32 & 30.23 & 10.01 & 83.98 & 78.96 & 82.21 & 82.34& 33.01\\
&$10^{-1}$&9.99&9.95&10.42&12.51&27.88&24.30&45.82&11.74&36.01&9.97&84.47&79.92&83.08&83.82&37.85\\
&$10^{-2}$&10.29&10.00&10.00&10.55&9.98&12.97&24.59&10.00&25.05&10.00&84.98&90.37&84.11&83.35&34.02\\
\hline\hline
\multirow{4}{*}{$\omega_2$} 
&10&9.97 & 10.02 & 10.09 & 11.33 & 11.17 & 11.27 & 31.21 & 10.55 & 24.64 & 10.20 & 84.68 & 85.82 & 84.51 & 83.61 & 34.65\\
% &1(DH)&10.00&10.00&11.25&10.02&10.59&10.04&13.53&10.00&10.00&10.00&72.97&23.62&10.00&{16.31}\\
&1(DH)& 10.08 & 10.00 & 10.81 & 10.05 & 10.20 & 15.58 & 17.31 & 10.32 & 30.23 & 10.01 & 83.98 & 78.96 & 82.21 & 82.34& 33.01\\
&$10^{-1}$& 10.45 & 12.71 &10.94&11.57&19.92 &17.37 &42.37&10.44&31.12&10.95&85.07&86.74&83.73&83.65 & 36.93\\
&$10^{-2}$&10.00&10.02&9.87&10.02&10.62&29.80&36.62&10.00&23.94&10.00&84.78&90.50&84.42&84.24 & 36.95\\
\hline\hline
\multirow{4}{*}{$\omega_3$} 
&$10^{-1}$ & 10.30 & 11.38 & 11.13 & 10.23 & 15.27 & 14.69 & 37.59 & 10.04 & 29.13 & 10.84 & 84.91 & 87.93 & 84.41 & 83.41 & 35.80\\
&$10^{-2}$ & 10.09 & 9.85 & 10.76 & 11.03 & 10.51 & 12.42 & 33.74 & 11.47 & 40.79 & 9.33 & 85.18 & 86.68 & 84.29 & 82.85 & 35.64\\
&$10^{-3}$ & 10.57 & 9.84 & 10.46 & 14.06 & 22.34 & 44.82 & 43.90 & 10.00 & 37.76 & 6.18 & 84.26 & 80.49 & 82.92 & 83.07 & 38.62\\
&$10^{-4}$(DH)& 10.08 & 10.00 & 10.81 & 10.05 & 10.20 & 15.58 & 17.31 & 10.32 & 30.23 & 10.01 & 83.98 & 78.96 & 82.21 & 82.34& 33.01\\
\hline 

\end{tabular}
}
\end{table*}

\subsubsection{Effectiveness of Semantic Images Generation (SIG) module}
To grasp the critical importance of SIG in our methodology, we undertake an ablation study on SIG including Text Prompts Clustering (TPC), and Stable Diffusion Model and ControlNet (SD+C).
The results are shown in Table~\ref{Tab:Ablation}. In the class-wise setting, we find that the improvement of the generation module is marginal. {However, in the sample-wise setting, the SIG can degrade the mean accuracy from 87.65\% to 37.47\%. }When we disentangle the TPC and SD+C, {we find that SD+C contributes the most and TPC contributes around 9\% reduction.} The ablation study shows that the SIG model in our proposed method plays an important role in sample-wise unlearnable examples to make the robust unlearnable examples. 

\begin{table*}[t]
\centering
\renewcommand\arraystretch{1.15}
\caption{{Evaluation on the different hiding model trained by solely controlling the hiding loss ($\mathcal{L}_{\text {hide}}$) and using our designed loss ($\mathcal{L}_{\text{total}}$). The test accuracy (\%) is evaluated on CIFAR-10 in the class-wise setting.}}
\label{Lhide}
\resizebox{1.0\linewidth}{!}{
\begin{tabular}{c|p{0.6cm}p{0.6cm}p{0.6cm}p{0.6cm}|p{0.6cm}p{0.7cm}p{0.6cm}p{0.6cm}p{0.6cm}p{0.6cm}p{0.6cm}p{0.8cm}|p{0.8cm}p{0.65cm}|c}
\hline
Method & Vanilla & Cutout  & Cutmix  & Mixup   & MeanF& MedianF & BDR & Gray & GaussN & GaussF & JPEG10 &JPEG50 & AT($L_{\infty}$) & AT($\ell_2$) & Mean\\
\hline
  Clean & 94.59 & 95.00   & 94.77 & 94.96  & 49.70   & 86.64   & 89.07   & 92.80  & 88.71   & 94.54   & 85.22 & 90.89& 84.19 & 83.54 & 87.07 \\

  Only $\mathcal{L}_{\text {hide}}$ & 94.69 & 94.95 & 94.48 & 94.86 & 56.45 & 86.10 & 89.06 & 92.79 & 88.49 & 94.44 & 85.24 & 90.75 & 84.07 & 83.46 & 87.89\\

DH & 10.08 & 10.00 & 10.81 & 10.05 & 10.20 & 15.58 & 17.31 & 10.32 & 30.23 & 10.01 & 83.98 & 78.96 & 82.21 & 82.34& 33.01 \\
\hline
\end{tabular}
}
\end{table*}

\begin{figure}[!t]
\begin{center}
\includegraphics[width=0.45\textwidth]{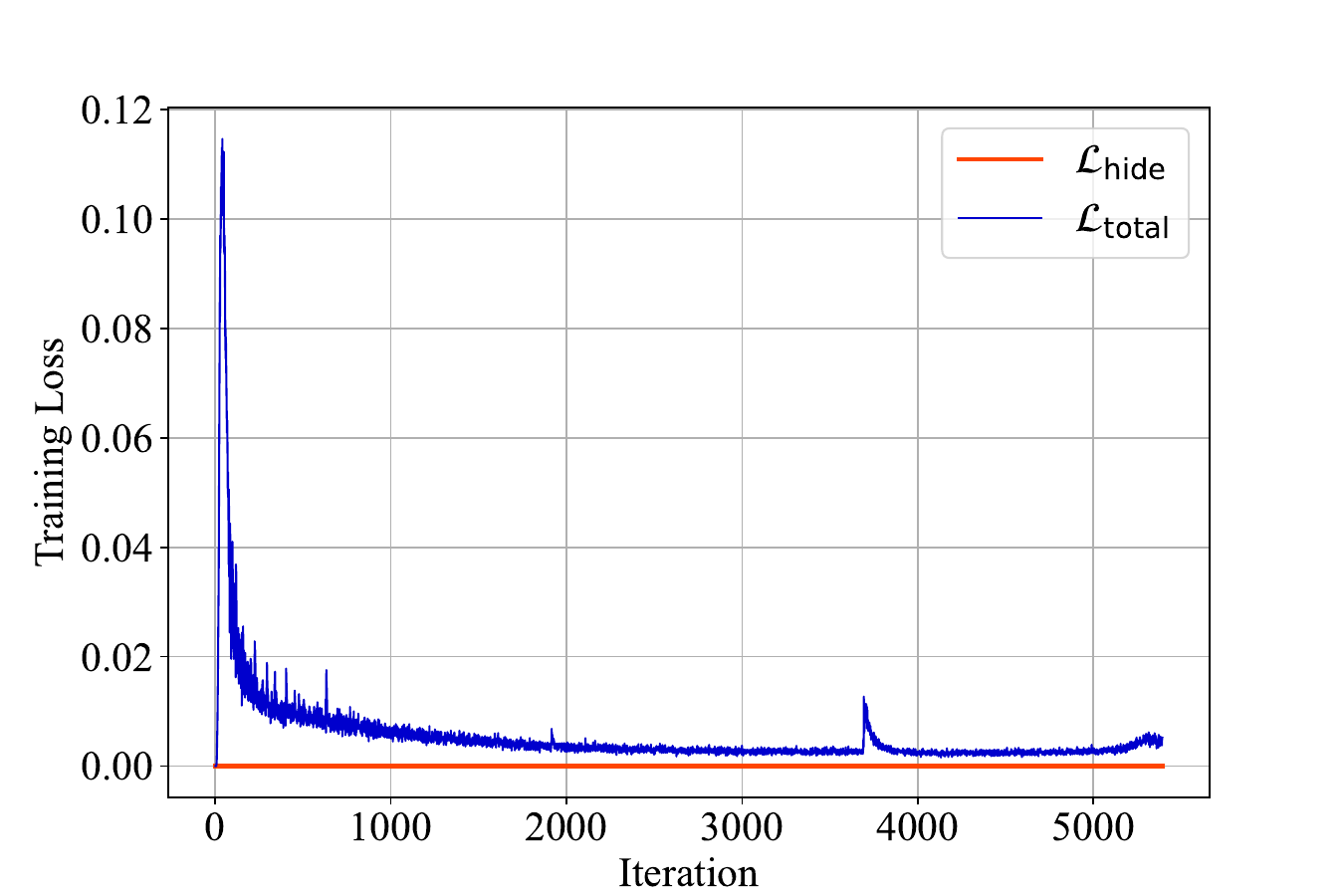}
\vspace{-1mm}
\end{center}
\caption{
Evaluating the dynamics of training loss: a comparative study using $\mathcal{L}_{\mathrm{hide}}$ and $\mathcal{L}_{\mathrm{total}}$ loss functions during the training process to observe the effectiveness of the revealing process in our proposed deep hiding model.
}
\label{Fig:8}
\end{figure}

\begin{table*}[t]
\centering
\renewcommand\arraystretch{1.15}
\caption{{Evaluation of the existing image hiding methods in other attacks. The test accuracy (\%) is evaluated on CIFAR-10.}}
\label{backdoor}
\resizebox{1.0\linewidth}{!}{
\begin{tabular}{c|p{0.6cm}p{0.6cm}p{0.6cm}p{0.6cm}|p{0.6cm}p{0.7cm}p{0.6cm}p{0.6cm}p{0.6cm}p{0.6cm}p{0.6cm}p{0.8cm}|p{0.8cm}p{0.65cm}|c}
\hline
  Method & Vanilla & Cutout  & Cutmix  & Mixup   & MeanF& MedianF & BDR & Gray & GaussN & GaussF & JPEG10 &JPEG50 &AT($L_{\infty}$) & AT($\ell_2$) & Mean\\

\hline
  Clean & 94.59   & 95.00   & 94.77 & 94.96  & 49.70   & 86.64   & 89.07   & 92.80  & 88.71   & 94.54   & 85.22 & 90.89 & 84.19 &83.54 & 87.47 \\
  Poison Ink~\cite{zhang2022poison} & 93.58 & 93.55 & 93.16 & 94.21 & 36.42 & 82.92 & 89.12 & 92.19 & 89.12 & 93.30 & 84.35 & 90.07 & 87.10 & 84.52 &85.97\\
  
  AdvINN~\cite{chen2023imperceptible} & 85.21 & 79.29 & 81.34	& 80.23	& 49.38	& 82.25	& 77.39	& 87.46	& 84.38	& 88.66	& 82.01	& 87.68	& 84.91	& 83.27	& 80.96\\
DH & 10.08 & 10.00 & 10.81 & 10.05 & 10.20 & 15.58 & 17.31 & 10.32 & 30.23 & 10.01 & 83.98 & 78.96 & 82.21 & 82.34& 33.01 \\
\hline
\end{tabular}
}
\end{table*}

\begin{table*}[t]
\centering
%\tabcolsep=0.25cm
%\scriptsize
\renewcommand\arraystretch{1.15}
\caption{{Test accuracy (\%) of models trained on unlearnable examples with a soft restriction setting against data augmentations, data preprocessing, and adversarial training. }}
\label{Soft_Restriction}
\resizebox{1\linewidth}{!}{
\begin{tabular}{c|c|p{0.6cm}p{0.6cm}p{0.6cm}p{0.6cm}|p{0.6cm}p{0.7cm}p{0.65cm}p{0.65cm}p{0.65cm}p{0.65cm}p{0.65cm}p{0.8cm}|cc|c|c}
\hline
 & Method & Vanilla & Cutout  & Cutmix  & Mixup   & MeanF& MedianF & BDR & Gray & GaussN & GaussF & JPEG10 &JPEG50 & AT($L_{\infty}$) & AT($\ell_2$) & Mean&PSNR\\
\hline
\multirow{4}{*}{\rotatebox{90}{CIFAR-10}} &DH(S) & 12.93 & 14.03 & 13.80 & 15.89 & 16.06 & 25.56 & 28.28 & 25.65 & 65.11 & 12.30 & 84.93 & 89.18 & 83.44 & 83.27& 37.47& 33.52\\

& DH'(S) & 15.36 & {10.79} & {10.00} & 14.72 & 17.68 & {17.00} & {21.12} & 17.61 & 22.78 & 11.16 & {80.41} & 81.03 &38.31 & 68.03 & 30.43& 31.63\\

\cline{2-18}

&DH(C) & {10.08} & {10.00} & {10.81} & {10.05} & {10.20} & {15.58} & {17.31} & {10.32} & 30.23 & {10.01} & 83.98 & 78.96 & 82.21 & 82.34& {33.01}& 33.55\\

&DH'(C) & {10.00} & {10.00} & {11.25} & {10.02} & {10.59} & {10.04} & {13.53} & {10.00} & {10.00} & {10.00} & {72.97} & {23.62} &{10.00} & 18.80 & 16.49 & 31.76\\

\hline
\hline

\multirow{4}{*}{\rotatebox{90}{CIFAR-100}} &DH(S) & 7.81 & 4.80 & 10.15 & 10.27 & 10.14 & 22.23 & 38.13 & 15.50 & 52.43 & 7.97 & 56.01& 65.72 & 56.48 & 56.38& 29.57&33.71\\

&DH'(S)& 4.79 & 4.13 & 5.39 & 4.72 & 6.22 & 10.21 & {12.12} & {3.72} & {19.85} & {3.61} & {49.50} & {34.86} & {41.12} & 37.56 & 16.99& 30.98\\

\cline{2-18}

&DH(C) & {1.22} & {1.01} & {1.22} & {1.09} & {1.51} & {2.72} & {12.08} & {0.96} & 19.86 & {1.01} & 55.07& 53.83 & 56.91 & 56.34& {18.91}& 33.69\\

&DH'(C) &{1.47} & {1.03} & {1.06} & {1.47} & {1.04} & {1.45} & {1.72} & {1.38} & {1.08} & {1.00} & {44.58} & {25.45} & {1.39} & 17.14 & 7.23& 31.26\\

\hline

\end{tabular}
}
\end{table*}

\subsubsection{Effectiveness of different hyper-parameters' settings} In our experiments, we have conducted some experiments of different hyper-parameters settings on CIFAR-10 and tabulated the results in class-wise setting, as shown in Table~\ref{Par_ablation}. The optimal results were obtained when $\omega_1$ and $\omega_2$ were set to 1, and $\omega_3$ was set to 0.0001. This configuration is effective because the loss associated with $\omega_3$ is typically 1000 times larger than those of $\omega_1$ and $\omega_2$. Setting $\omega_3$ to 0.0001 helps balance the training process by ensuring that the scales of the losses remain consistent. 

Besides, to observe the effectiveness of the revealing process in our proposed deep hiding model, we test the performance when $\omega_1$, $\omega_2$, and $\omega_3$ are set as 0, which means we just focus exclusively on the hiding loss, denoted as $\mathcal{L}_{\text {hide}}$. The training loss trends for using only $\mathcal{L}_{\text {hide}}$ compared to our complete loss $\mathcal{L}_{\text{total}}$ are depicted in Fig. \ref{Fig:8}. The $\mathcal{L}_{\text{total}}$ exhibits a notable initial rise and subsequent decline, attributed to the optimization of the revealing loss. In contrast, when training with only the hiding loss, the training plot shows the loss remains 0. Since the minimal amount of information is hidden in the clean images during the first step, with the perturbation radius staying below the $8/255$ threshold, fulfilling the optimization objectives without further optimization is needed. 
Furthermore, our evaluation of unlearnable examples generated by the model trained only with $\mathcal{L}_{hide}$ reveals that the test accuracy is close to that of clean images, as shown in Table~\ref{Lhide}. This indicates that minimal information is hidden in clean images, leading to ineffective unlearnability. Based on the above experimental results and our analysis, we apply the designed total loss $\mathcal{L}_{\text{total}}$ in all experiments.

\subsection{Evaluation of existing image hiding methods in other attacks}
{Existing image hiding methods have been applied in other forms of attacks, such as Poison Ink~\cite{zhang2022poison} for backdoor attack, and AdvINN~\cite{chen2023imperceptible} for adversarial attack. In this part, we investigate whether these methods could also generate unlearnable examples. The experimental results are shown in Table~\ref{backdoor}. 
Based on the experimental results, the vanilla test accuracies of Poison Ink and AdvINN are 93.58 and 85.21, respectively.
Such results indicate that they do not suit the goal of unlearnable examples generation. The adversarial noise added by Poison Ink is based on the contour features of the clean image, which are strongly correlated with the clean image itself, making it difficult to achieve unlearnability. 
AdvINN adaptively generates a target image that starts with a constant image (\eg,  all pixels set to 0.5), and the resulting pattern remains strongly linked to clean images. The purpose of these attacks differs from that of creating unlearnable examples. Although hiding technology is utilized in this field to create imperceptible samples, it cannot be directly employed to achieve unlearnability.}

\subsection{Performance in a soft restriction setting}
{
In this section, we investigate the balance between imperceptibility and protection performance by using a soft restriction strategy. Specifically, we utilize our proposed method to generate unlearnable examples without applying clipping operations to restrict the perturbation within $8/255$. 
It is designed upon the principle of invisibility, which focuses on perturbing regions less discernible to humans. Similar strategies have been employed in~\cite{Sadasivan_2023_CVPR, ahn2024imperceptible} to prioritize imperceptibility instead of bound restriction.
The experimental results are presented in Table~\ref{Soft_Restriction}.
It is important to note that even without the bound restriction for our generated unlearnable examples, the invisibility of our examples can be preserved due to the adaptive hiding manner employed by our hiding model. 
This manner allows for the concealment of perturbations of varying scales across different regions. 
In a soft restriction setting, our method significantly enhances the robustness of our unlearnable examples, especially showing a good performance against AT. 
This phenomenon indicates the general robustness of our method, underscoring its potential practicality in real-world scenarios.}

\section{Conclusion}
{In this paper, we present a novel Deep Hiding scheme tailored for the generation of robust unlearnable examples.}
By embedding clean images with semantically rich high-level attributes, we ensure that the generated unlearnable examples effectively derail the learning processes of unauthorized deep learning models.
Additionally, our uniquely conceived Latent Feature Concentration (LFC) module further enhances the effectiveness of unlearnable examples by regularizing the intra-class variance of perturbations.
To guarantee the robustness of unlearnable examples, we introduce the Semantic Images Generation (SIG) module to generate hidden semantic images by maintaining semantic feature consistency within each class.
The extensive experimental results demonstrate that our proposed method achieves outstanding unlearnability performance and superior robustness against {various} countermeasures.

\bibliographystyle{IEEEtran}
\bibliography{reference, IEEEabrv}

\begin{thebibliography}{10}
\providecommand{\url}[1]{#1}
\csname url@samestyle\endcsname
\providecommand{\newblock}{\relax}
\providecommand{\bibinfo}[2]{#2}
\providecommand{\BIBentrySTDinterwordspacing}{\spaceskip=0pt\relax}
\providecommand{\BIBentryALTinterwordstretchfactor}{4}
\providecommand{\BIBentryALTinterwordspacing}{\spaceskip=\fontdimen2\font plus
\BIBentryALTinterwordstretchfactor\fontdimen3\font minus \fontdimen4\font\relax}
\providecommand{\BIBforeignlanguage}[2]{{%
\expandafter\ifx\csname l@#1\endcsname\relax
\typeout{** WARNING: IEEEtran.bst: No hyphenation pattern has been}%
\typeout{** loaded for the language `#1'. Using the pattern for}%
\typeout{** the default language instead.}%
\else
\language=\csname l@#1\endcsname
\fi
#2}}
\providecommand{\BIBdecl}{\relax}
\BIBdecl

\bibitem{intro1}
D.~Mahajan, R.~Girshick, V.~Ramanathan, K.~He, M.~Paluri, Y.~Li, A.~Bharambe, and L.~Van Der~Maaten, ``Exploring the limits of weakly supervised pretraining,'' in \emph{Proceedings of the European conference on computer vision (ECCV)}, 2018, pp. 181--196.

\bibitem{intro2}
V.~U. Prabhu and A.~Birhane, ``Large image datasets: A pyrrhic win for computer vision?'' \emph{arXiv preprint arXiv:2006.16923}, 2020.

\bibitem{privacy}
H.~Drachsler and W.~Greller, ``Privacy and analytics: it's a delicate issue a checklist for trusted learning analytics,'' in \emph{Proceedings of the sixth international conference on learning analytics \& knowledge}, 2016, pp. 89--98.

\bibitem{TC}
J.~Shen, X.~Zhu, and D.~Ma, ``Tensorclog: An imperceptible poisoning attack on deep neural network applications,'' \emph{IEEE Access}, vol.~7, pp. 41\,498--41\,506, 2019.

\bibitem{DC}
J.~Feng, Q.-Z. Cai, and Z.-H. Zhou, ``Learning to confuse: generating training time adversarial data with auto-encoder,'' \emph{Advances in Neural Information Processing Systems}, vol.~32, 2019.

\bibitem{EM}
H.~Huang, X.~Ma, S.~M. Erfani, J.~Bailey, and Y.~Wang, ``Unlearnable examples: Making personal data unexploitable,'' \emph{arXiv preprint arXiv:2101.04898}, 2021.

\bibitem{HYPO}
L.~Tao, L.~Feng, J.~Yi, S.-J. Huang, and S.~Chen, ``Better safe than sorry: Preventing delusive adversaries with adversarial training,'' \emph{Advances in Neural Information Processing Systems}, vol.~34, pp. 16\,209--16\,225, 2021.

\bibitem{TAP}
L.~Fowl, M.~Goldblum, P.-y. Chiang, J.~Geiping, W.~Czaja, and T.~Goldstein, ``Adversarial examples make strong poisons,'' \emph{Advances in Neural Information Processing Systems}, vol.~34, pp. 30\,339--30\,351, 2021.

\bibitem{lin2024safeguarding}
X.~Lin, Y.~Yu, S.~Xia, J.~Jiang, H.~Wang, Z.~Yu, Y.~Liu, Y.~Fu, S.~Wang, W.~Tang \emph{et~al.}, ``Safeguarding medical image segmentation datasets against unauthorized training via contour-and texture-aware perturbations,'' \emph{arXiv preprint arXiv:2403.14250}, 2024.

\bibitem{radiya2021data}
E.~Radiya-Dixit, S.~Hong, N.~Carlini, and F.~Tramer, ``Data poisoning won’t save you from facial recognition,'' in \emph{International Conference on Learning Representations}, 2021.

\bibitem{REM}
S.~Fu, F.~He, Y.~Liu, L.~Shen, and D.~Tao, ``Robust unlearnable examples: Protecting data privacy against adversarial learning,'' in \emph{International Conference on Learning Representations}, 2021.

\bibitem{ADVIN}
Z.~Wang, Y.~Wang, and Y.~Wang, ``Fooling adversarial training with inducing noise,'' \emph{arXiv preprint arXiv:2111.10130}, 2021.

\bibitem{EntF}
R.~Wen, Z.~Zhao, Z.~Liu, M.~Backes, T.~Wang, and Y.~Zhang, ``Is adversarial training really a silver bullet for mitigating data poisoning?'' in \emph{The Eleventh International Conference on Learning Representations}, 2022.

\bibitem{madry2017towards}
A.~Madry, A.~Makelov, L.~Schmidt, D.~Tsipras, and A.~Vladu, ``Towards deep learning models resistant to adversarial attacks,'' \emph{arXiv preprint arXiv:1706.06083}, 2017.

\bibitem{yu2022towards}
Y.~Yu, W.~Yang, Y.-P. Tan, and A.~C. Kot, ``Towards robust rain removal against adversarial attacks: A comprehensive benchmark analysis and beyond,'' in \emph{Proceedings of the IEEE/CVF Conference on Computer Vision and Pattern Recognition}, 2022, pp. 6013--6022.

\bibitem{xia2024mitigating}
\BIBentryALTinterwordspacing
S.~Xia, Y.~Yu, X.~Jiang, and H.~Ding, ``Mitigating the curse of dimensionality for certified robustness via dual randomized smoothing,'' in \emph{The Twelfth International Conference on Learning Representations}, 2024. [Online]. Available: \url{https://openreview.net/forum?id=C1sQBG6Sqp}
\BIBentrySTDinterwordspacing

\bibitem{ISS}
Z.~Liu, Z.~Zhao, and M.~A. Larson, ``Image shortcut squeezing: Countering perturbative availability poisons with compression,'' in \emph{International Conference on Machine Learning, {ICML} 2023, 23-29 July 2023, Honolulu, Hawaii, {USA}}, ser. Proceedings of Machine Learning Research, vol. 202, 2023, pp. 22\,473--22\,487.

\bibitem{OPS}
S.~Wu, S.~Chen, C.~Xie, and X.~Huang, ``One-pixel shortcut: on the learning preference of deep neural networks,'' \emph{arXiv preprint arXiv:2205.12141}, 2022.

\bibitem{shape}
R.~Geirhos, P.~Rubisch, C.~Michaelis, M.~Bethge, F.~A. Wichmann, and W.~Brendel, ``Imagenet-trained cnns are biased towards texture; increasing shape bias improves accuracy and robustness,'' \emph{arXiv preprint arXiv:1811.12231}, 2018.

\bibitem{high1}
M.~D. Zeiler and R.~Fergus, ``Visualizing and understanding convolutional networks,'' in \emph{Computer Vision--ECCV 2014: 13th European Conference, Zurich, Switzerland, September 6-12, 2014, Proceedings, Part I 13}.\hskip 1em plus 0.5em minus 0.4em\relax Springer, 2014, pp. 818--833.

\bibitem{high2}
K.~He, X.~Zhang, S.~Ren, and J.~Sun, ``Deep residual learning for image recognition,'' in \emph{Proceedings of the IEEE conference on computer vision and pattern recognition}, 2016, pp. 770--778.

\bibitem{zhou2023generative}
Z.~Zhou, X.~Dong, R.~Meng, M.~Wang, H.~Yan, K.~Yu, and K.-K.~R. Choo, ``Generative steganography via auto-generation of semantic object contours,'' \emph{IEEE Transactions on Information Forensics and Security}, 2023.

\bibitem{high3}
X.~Li, Y.~Grandvalet, F.~Davoine, J.~Cheng, Y.~Cui, H.~Zhang, S.~Belongie, Y.-H. Tsai, and M.-H. Yang, ``Transfer learning in computer vision tasks: Remember where you come from,'' \emph{Image and Vision Computing}, vol.~93, p. 103853, 2020.

\bibitem{baluja2017hiding}
S.~Baluja, ``Hiding images in plain sight: Deep steganography,'' \emph{Advances in neural information processing systems}, vol.~30, 2017.

\bibitem{yu2020attention}
C.~Yu, ``Attention based data hiding with generative adversarial networks,'' in \emph{Proceedings of the AAAI conference on artificial intelligence}, vol.~34, no.~01, 2020, pp. 1120--1128.

\bibitem{Hinet}
J.~Jing, X.~Deng, M.~Xu, J.~Wang, and Z.~Guan, ``Hinet: deep image hiding by invertible network,'' in \emph{Proceedings of the IEEE/CVF international conference on computer vision}, 2021, pp. 4733--4742.

\bibitem{zhang2019invisible}
R.~Zhang, S.~Dong, and J.~Liu, ``Invisible steganography via generative adversarial networks,'' \emph{Multimedia tools and applications}, vol.~78, pp. 8559--8575, 2019.

\bibitem{tang2020automatic}
W.~Tang, B.~Li, M.~Barni, J.~Li, and J.~Huang, ``An automatic cost learning framework for image steganography using deep reinforcement learning,'' \emph{IEEE Transactions on Information Forensics and Security}, vol.~16, pp. 952--967, 2020.

\bibitem{pan2021seek}
W.~Pan, Y.~Yin, X.~Wang, Y.~Jing, and M.~Song, ``Seek-and-hide: adversarial steganography via deep reinforcement learning,'' \emph{IEEE Transactions on Pattern Analysis and Machine Intelligence}, vol.~44, no.~11, pp. 7871--7884, 2021.

\bibitem{cui2024meta}
Q.~Cui, W.~Tang, Z.~Zhou, R.~Meng, G.~Nan, and Y.-Q. Shi, ``Meta security metric learning for secure deep image hiding,'' \emph{IEEE Transactions on Dependable and Secure Computing}, 2024.

\bibitem{hu2023invisible}
X.~Hu, Z.~Fu, X.~Zhang, and Y.~Chen, ``Invisible and steganalysis-resistant deep image hiding based on one-way adversarial invertible networks,'' \emph{IEEE Transactions on Circuits and Systems for Video Technology}, 2023.

\bibitem{deepmih}
Z.~Guan, J.~Jing, X.~Deng, M.~Xu, L.~Jiang, Z.~Zhang, and Y.~Li, ``Deepmih: Deep invertible network for multiple image hiding,'' \emph{IEEE Transactions on Pattern Analysis and Machine Intelligence}, vol.~45, no.~1, pp. 372--390, 2022.

\bibitem{INN-based}
M.~Xiao, S.~Zheng, C.~Liu, Z.~Lin, and T.-Y. Liu, ``Invertible rescaling network and its extensions,'' \emph{International Journal of Computer Vision}, vol. 131, no.~1, pp. 134--159, 2023.

\bibitem{meng2022traceable}
R.~Meng, Z.~Zhou, Q.~Cui, K.-Y. Lam, and A.~Kot, ``Traceable and authenticable image tagging for fake news detection,'' \emph{arXiv preprint arXiv:2211.10923}, 2022.

\bibitem{Traceble}
------, ``Traceable and authenticable image tagging for fake news detection,'' \emph{arXiv preprint arXiv:2211.10923}, 2022.

\bibitem{resnet18}
K.~He, X.~Zhang, S.~Ren, and J.~Sun, ``Deep residual learning for image recognition,'' in \emph{Proceedings of the IEEE conference on computer vision and pattern recognition}, 2016, pp. 770--778.

\bibitem{EOT}
A.~Athalye, L.~Engstrom, A.~Ilyas, and K.~Kwok, ``Synthesizing robust adversarial examples,'' in \emph{International conference on machine learning}.\hskip 1em plus 0.5em minus 0.4em\relax PMLR, 2018, pp. 284--293.

\bibitem{TUE}
J.~Ren, H.~Xu, Y.~Wan, X.~Ma, L.~Sun, and J.~Tang, ``Transferable unlearnable examples,'' \emph{arXiv preprint arXiv:2210.10114}, 2022.

\bibitem{zhang2023unlearnable}
J.~Zhang, X.~Ma, Q.~Yi, J.~Sang, Y.-G. Jiang, Y.~Wang, and C.~Xu, ``Unlearnable clusters: Towards label-agnostic unlearnable examples,'' in \emph{Proceedings of the IEEE/CVF Conference on Computer Vision and Pattern Recognition}, 2023, pp. 3984--3993.

\bibitem{AR}
P.~Sandoval-Segura, V.~Singla, J.~Geiping, M.~Goldblum, T.~Goldstein, and D.~Jacobs, ``Autoregressive perturbations for data poisoning,'' \emph{Advances in Neural Information Processing Systems}, vol.~35, pp. 27\,374--27\,386, 2022.

\bibitem{LSP}
D.~Yu, H.~Zhang, W.~Chen, J.~Yin, and T.-Y. Liu, ``Availability attacks create shortcuts,'' in \emph{Proceedings of the 28th ACM SIGKDD Conference on Knowledge Discovery and Data Mining}, 2022, pp. 2367--2376.

\bibitem{zhang2023apmsa}
J.~Zhang, S.~Peng, Y.~Gao, Z.~Zhang, and Q.~Hong, ``Apmsa: adversarial perturbation against model stealing attacks,'' \emph{IEEE Transactions on Information Forensics and Security}, vol.~18, pp. 1667--1679, 2023.

\bibitem{xu2022robust}
Y.~Xu, C.~Mou, Y.~Hu, J.~Xie, and J.~Zhang, ``Robust invertible image steganography,'' in \emph{Proceedings of the IEEE/CVF Conference on Computer Vision and Pattern Recognition}, 2022, pp. 7875--7884.

\bibitem{zhu2023information}
Y.~Zhu, Y.~Chen, X.~Li, R.~Zhang, X.~Tian, B.~Zheng, and Y.~Chen, ``Information-containing adversarial perturbation for combating facial manipulation systems,'' \emph{IEEE Transactions on Information Forensics and Security}, vol.~18, pp. 2046--2059, 2023.

\bibitem{lee2023robust}
H.~J. Lee and Y.~M. Ro, ``Robust proxy: Improving adversarial robustness by robust proxy learning,'' \emph{IEEE Transactions on Information Forensics and Security}, 2023.

\bibitem{cutout}
T.~DeVries and G.~W. Taylor, ``Improved regularization of convolutional neural networks with cutout,'' \emph{arXiv preprint arXiv:1708.04552}, 2017.

\bibitem{cutmix}
S.~Yun, D.~Han, S.~J. Oh, S.~Chun, J.~Choe, and Y.~Yoo, ``Cutmix: Regularization strategy to train strong classifiers with localizable features,'' in \emph{Proceedings of the IEEE/CVF international conference on computer vision}, 2019, pp. 6023--6032.

\bibitem{mixup}
H.~Zhang, M.~Cisse, Y.~N. Dauphin, and D.~Lopez-Paz, ``mixup: Beyond empirical risk minimization,'' \emph{arXiv preprint arXiv:1710.09412}, 2017.

\bibitem{qin2023learning}
T.~Qin, X.~Gao, J.~Zhao, K.~Ye, and C.-Z. Xu, ``Learning the unlearnable: Adversarial augmentations suppress unlearnable example attacks,'' \emph{arXiv preprint arXiv:2303.15127}, 2023.

\bibitem{AVATAR}
H.~M. Dolatabadi, S.~Erfani, and C.~Leckie, ``The devil's advocate: Shattering the illusion of unexploitable data using diffusion models,'' \emph{arXiv preprint arXiv:2303.08500}, 2023.

\bibitem{nie2022diffusion}
W.~Nie, B.~Guo, Y.~Huang, C.~Xiao, A.~Vahdat, and A.~Anandkumar, ``Diffusion models for adversarial purification,'' \emph{arXiv preprint arXiv:2205.07460}, 2022.

\bibitem{sandoval2024can}
P.~Sandoval-Segura, V.~Singla, J.~Geiping, M.~Goldblum, and T.~Goldstein, ``What can we learn from unlearnable datasets?'' \emph{Advances in Neural Information Processing Systems}, vol.~36, 2024.

\bibitem{yu2024purify}
Y.~Yu, Y.~Wang, S.~Xia, W.~Yang, S.~Lu, Y.-P. Tan, and A.~C. Kot, ``Purify unlearnable examples via rate-constrained variational autoencoders,'' in \emph{International Conference on Machine Learning, {ICML} 2024}, ser. Proceedings of Machine Learning Research, 2024.

\bibitem{zhao2021invertible}
R.~Zhao, T.~Liu, J.~Xiao, D.~P. Lun, and K.-M. Lam, ``Invertible image decolorization,'' \emph{IEEE Transactions on Image Processing}, vol.~30, pp. 6081--6095, 2021.

\bibitem{huang2022winnet}
J.-J. Huang and P.~L. Dragotti, ``Winnet: Wavelet-inspired invertible network for image denoising,'' \emph{IEEE Transactions on Image Processing}, vol.~31, pp. 4377--4392, 2022.

\bibitem{jing2021hinet}
J.~Jing, X.~Deng, M.~Xu, J.~Wang, and Z.~Guan, ``Hinet: deep image hiding by invertible network,'' in \emph{Proceedings of the IEEE/CVF international conference on computer vision}, 2021, pp. 4733--4742.

\bibitem{guan2022deepmih}
Z.~Guan, J.~Jing, X.~Deng, M.~Xu, L.~Jiang, Z.~Zhang, and Y.~Li, ``Deepmih: Deep invertible network for multiple image hiding,'' \emph{IEEE Transactions on Pattern Analysis and Machine Intelligence}, vol.~45, no.~1, pp. 372--390, 2022.

\bibitem{li2023iscmis}
F.~Li, Y.~Sheng, X.~Zhang, and C.~Qin, ``iscmis: Spatial-channel attention based deep invertible network for multi-image steganography,'' \emph{IEEE Transactions on Multimedia}, 2023.

\bibitem{lu2021large}
S.-P. Lu, R.~Wang, T.~Zhong, and P.~L. Rosin, ``Large-capacity image steganography based on invertible neural networks,'' in \emph{Proceedings of the IEEE/CVF conference on computer vision and pattern recognition}, 2021, pp. 10\,816--10\,825.

\bibitem{zhong2020backdoor}
H.~Zhong, C.~Liao, A.~C. Squicciarini, S.~Zhu, and D.~Miller, ``Backdoor embedding in convolutional neural network models via invisible perturbation,'' in \emph{Proceedings of the Tenth ACM Conference on Data and Application Security and Privacy}, 2020, pp. 97--108.

\bibitem{zhang2022poison}
J.~Zhang, C.~Dongdong, Q.~Huang, J.~Liao, W.~Zhang, H.~Feng, G.~Hua, and N.~Yu, ``Poison ink: Robust and invisible backdoor attack,'' \emph{IEEE Transactions on Image Processing}, vol.~31, pp. 5691--5705, 2022.

\bibitem{chen2023imperceptible}
Z.~Chen, Z.~Wang, J.-J. Huang, W.~Zhao, X.~Liu, and D.~Guan, ``Imperceptible adversarial attack via invertible neural networks,'' in \emph{Proceedings of the AAAI Conference on Artificial Intelligence}, vol.~37, no.~1, 2023, pp. 414--424.

\bibitem{yu2023backdoor}
Y.~Yu, Y.~Wang, W.~Yang, S.~Lu, Y.-P. Tan, and A.~C. Kot, ``Backdoor attacks against deep image compression via adaptive frequency trigger,'' in \emph{Proceedings of the IEEE/CVF Conference on Computer Vision and Pattern Recognition}, 2023, pp. 12\,250--12\,259.

\bibitem{ldm}
R.~Rombach, A.~Blattmann, D.~Lorenz, P.~Esser, and B.~Ommer, ``High-resolution image synthesis with latent diffusion models,'' in \emph{Proceedings of the IEEE/CVF conference on computer vision and pattern recognition}, 2022, pp. 10\,684--10\,695.

\bibitem{controlnet}
L.~Zhang and M.~Agrawala, ``Adding conditional control to text-to-image diffusion models,'' \emph{arXiv preprint arXiv:2302.05543}, 2023.

\bibitem{prompt}
R.~Gandikota, J.~Materzy\'nska, J.~Fiotto-Kaufman, and D.~Bau, ``Erasing concepts from diffusion models,'' in \emph{Proceedings of the 2023 IEEE International Conference on Computer Vision}, 2023.

\bibitem{coco}
T.-Y. Lin, M.~Maire, S.~Belongie, J.~Hays, P.~Perona, D.~Ramanan, P.~Doll{\'a}r, and C.~L. Zitnick, ``Microsoft coco: Common objects in context,'' in \emph{Computer Vision--ECCV 2014: 13th European Conference, Zurich, Switzerland, September 6-12, 2014, Proceedings, Part V 13}.\hskip 1em plus 0.5em minus 0.4em\relax Springer, 2014, pp. 740--755.

\bibitem{kmeans}
D.~Arthur and S.~Vassilvitskii, ``K-means++ the advantages of careful seeding,'' in \emph{Proceedings of the eighteenth annual ACM-SIAM symposium on Discrete algorithms}, 2007, pp. 1027--1035.

\bibitem{clip}
A.~Radford, J.~W. Kim, C.~Hallacy, A.~Ramesh, G.~Goh, S.~Agarwal, G.~Sastry, A.~Askell, P.~Mishkin, J.~Clark \emph{et~al.}, ``Learning transferable visual models from natural language supervision,'' in \emph{International conference on machine learning}.\hskip 1em plus 0.5em minus 0.4em\relax PMLR, 2021, pp. 8748--8763.

\bibitem{zhang2023adding}
L.~Zhang, A.~Rao, and M.~Agrawala, ``Adding conditional control to text-to-image diffusion models,'' in \emph{Proceedings of the IEEE/CVF International Conference on Computer Vision}, 2023, pp. 3836--3847.

\bibitem{rombach2022high}
R.~Rombach, A.~Blattmann, D.~Lorenz, P.~Esser, and B.~Ommer, ``High-resolution image synthesis with latent diffusion models,'' in \emph{Proceedings of the IEEE/CVF conference on computer vision and pattern recognition}, 2022, pp. 10\,684--10\,695.

\bibitem{geirhos2020shortcut}
R.~Geirhos, J.-H. Jacobsen, C.~Michaelis, R.~Zemel, W.~Brendel, M.~Bethge, and F.~A. Wichmann, ``Shortcut learning in deep neural networks,'' \emph{Nature Machine Intelligence}, vol.~2, no.~11, pp. 665--673, 2020.

\bibitem{zipf2016human}
G.~K. Zipf, \emph{Human behavior and the principle of least effort: An introduction to human ecology}.\hskip 1em plus 0.5em minus 0.4em\relax Ravenio Books, 2016.

\bibitem{grad}
R.~R. Selvaraju, M.~Cogswell, A.~Das, R.~Vedantam, D.~Parikh, and D.~Batra, ``Grad-cam: Visual explanations from deep networks via gradient-based localization,'' in \emph{Proceedings of the IEEE international conference on computer vision}, 2017, pp. 618--626.

\bibitem{cifar10}
A.~Krizhevsky, G.~Hinton \emph{et~al.}, ``Learning multiple layers of features from tiny images,'' 2009.

\bibitem{imagenet}
J.~Deng, W.~Dong, R.~Socher, L.-J. Li, K.~Li, and L.~Fei-Fei, ``Imagenet: A large-scale hierarchical image database,'' in \emph{2009 IEEE conference on computer vision and pattern recognition}.\hskip 1em plus 0.5em minus 0.4em\relax Ieee, 2009, pp. 248--255.

\bibitem{VGG}
K.~Simonyan and A.~Zisserman, ``Very deep convolutional networks for large-scale image recognition,'' \emph{arXiv preprint arXiv:1409.1556}, 2014.

\bibitem{densenet}
G.~Huang, Z.~Liu, L.~Van Der~Maaten, and K.~Q. Weinberger, ``Densely connected convolutional networks,'' in \emph{Proceedings of the IEEE conference on computer vision and pattern recognition}, 2017, pp. 4700--4708.

\bibitem{VIT}
A.~Dosovitskiy, L.~Beyer, A.~Kolesnikov, D.~Weissenborn, X.~Zhai, T.~Unterthiner, M.~Dehghani, M.~Minderer, G.~Heigold, S.~Gelly \emph{et~al.}, ``An image is worth 16x16 words: Transformers for image recognition at scale,'' \emph{arXiv preprint arXiv:2010.11929}, 2020.

\bibitem{adam}
D.~P. Kingma and J.~Ba, ``Adam: A method for stochastic optimization,'' \emph{arXiv preprint arXiv:1412.6980}, 2014.

\bibitem{he2022indiscriminate}
H.~He, K.~Zha, and D.~Katabi, ``Indiscriminate poisoning attacks on unsupervised contrastive learning,'' \emph{arXiv preprint arXiv:2202.11202}, 2022.

\bibitem{Sadasivan_2023_CVPR}
V.~S. Sadasivan, M.~Soltanolkotabi, and S.~Feizi, ``Cuda: Convolution-based unlearnable datasets,'' in \emph{Proceedings of the IEEE/CVF Conference on Computer Vision and Pattern Recognition (CVPR)}, June 2023, pp. 3862--3871.

\bibitem{ahn2024imperceptible}
N.~Ahn, W.~Ahn, K.~Yoo, D.~Kim, and S.-H. Nam, ``Imperceptible protection against style imitation from diffusion models,'' \emph{arXiv preprint arXiv:2403.19254}, 2024.

\end{thebibliography}

%\newpage

\begin{IEEEbiography}[{\includegraphics[width=1in,height=1.25in,clip,keepaspectratio]{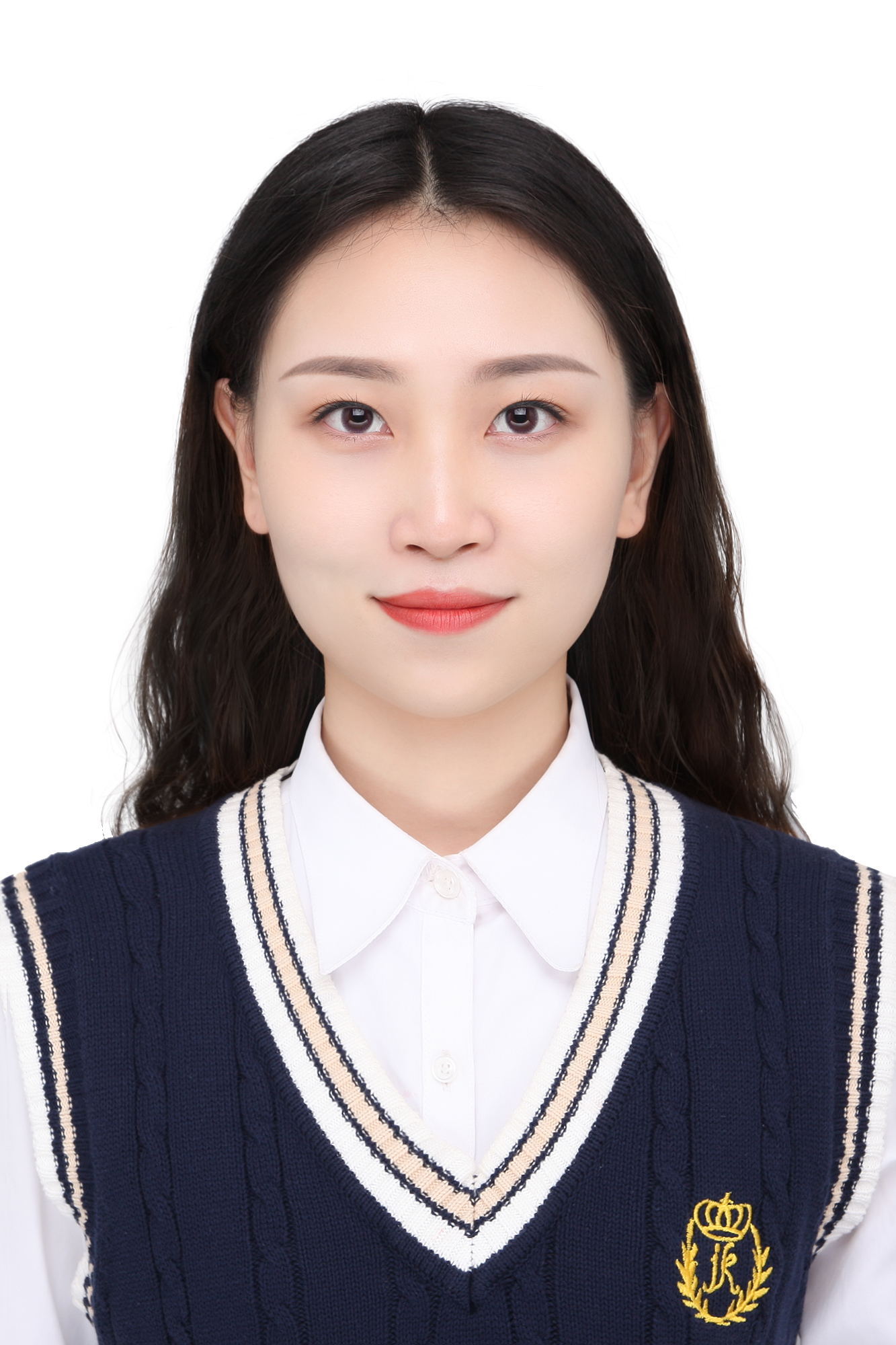}}]{Ruohan Meng} %(Member, IEEE) 
received her B.E. degree in Software Engineering and Ph.D. degree in Information Security from Nanjing University of Information Science and Technology, China in 2018 and 2023. She is currently a Research Fellow with the School of Electrical and Electronic Engineering, Nanyang Technological University, Singapore. Her research interests include information hiding, covert communication, and privacy protection.
\end{IEEEbiography}

\vspace{-1.2cm}

\begin{IEEEbiography}[{\includegraphics[width=1in,height=1.25in,clip,keepaspectratio]{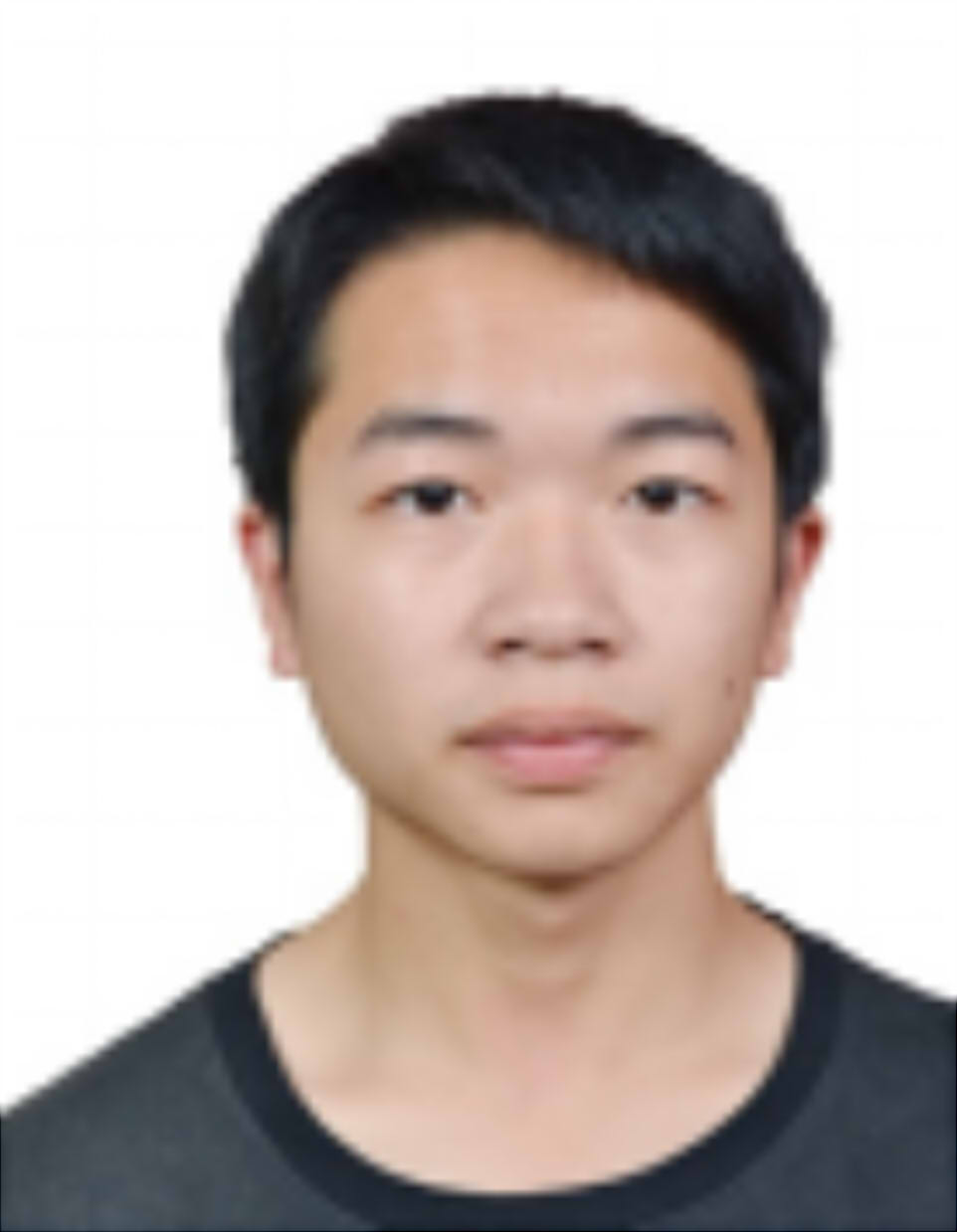}}]{ Chenyu Yi} received the BEng degree from Nanyang Technological University. He is currently pursuing the Ph.D. degree at Nanyang Technological University. His research interests include robust and trustworthy deep learning and visual understanding.  
\end{IEEEbiography}

\vspace{-1.1cm}
\begin{IEEEbiography}[{\includegraphics[width=1in,height=1.25in,clip,keepaspectratio]{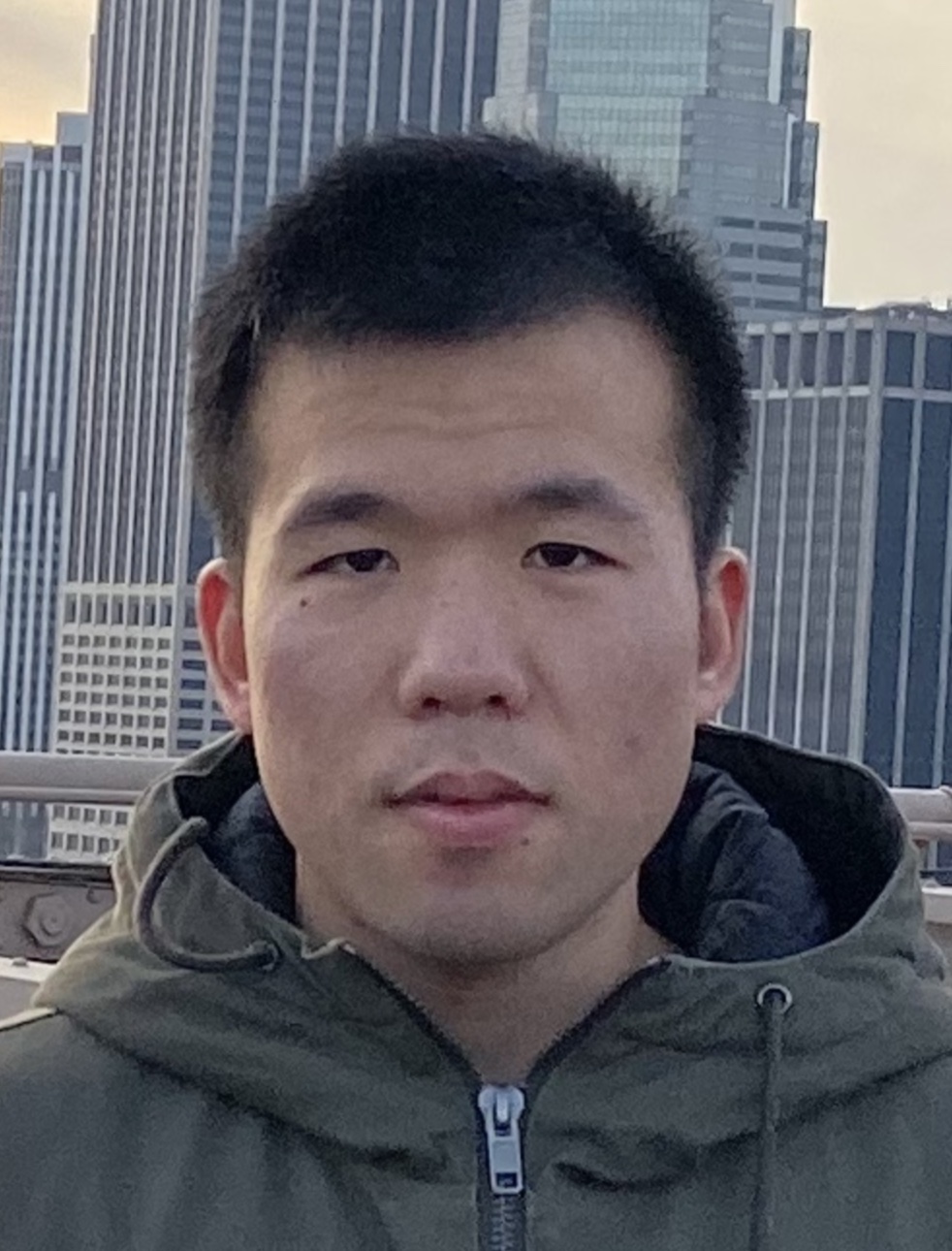}}]
{Yi Yu} 
received the B.S. degree from the Department of Automation, Tsinghua University, China and the M.S, degree from the Department of Electrical and Computer Engineering, University of California San Diego, United States, in 2019 and 2021, respectively. He is currently working toward the PhD degree in the ROSE Lab, Interdisciplinary Graduate Programme, Nanyang Technology University, Singapore.
His research interests include adversarial robustness and low-level vision.  
% He serves as a regular reviewer member for IEEE TNNLS, IEEE TCSVT, Neurips and CVPR.
\end{IEEEbiography}

\vspace{-1.2cm}
\begin{IEEEbiography}[{\includegraphics[width=1in,height=1.25in,clip,keepaspectratio]{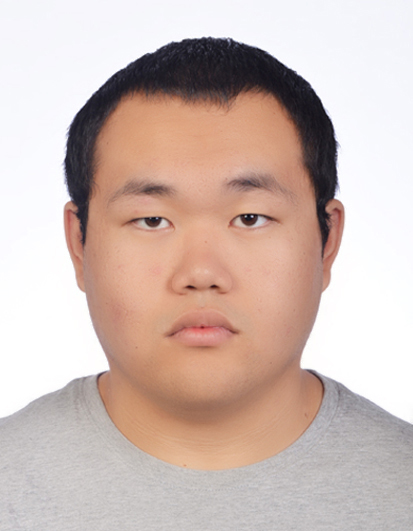}}]{Siyuan Yang} received the BEng degree from Harbin Institute of Technology and the MSc degree from Nanyang Technological University. He is currently pursuing the Ph.D. degree with the Interdisciplinary Graduate Programme, Nanyang Technological University. His research interests include computer vision, action recognition, and human pose estimation.  
\end{IEEEbiography}

\vspace{-1.2cm}
\begin{IEEEbiography}[{\includegraphics[width=1in,height=1.25in,clip,keepaspectratio]{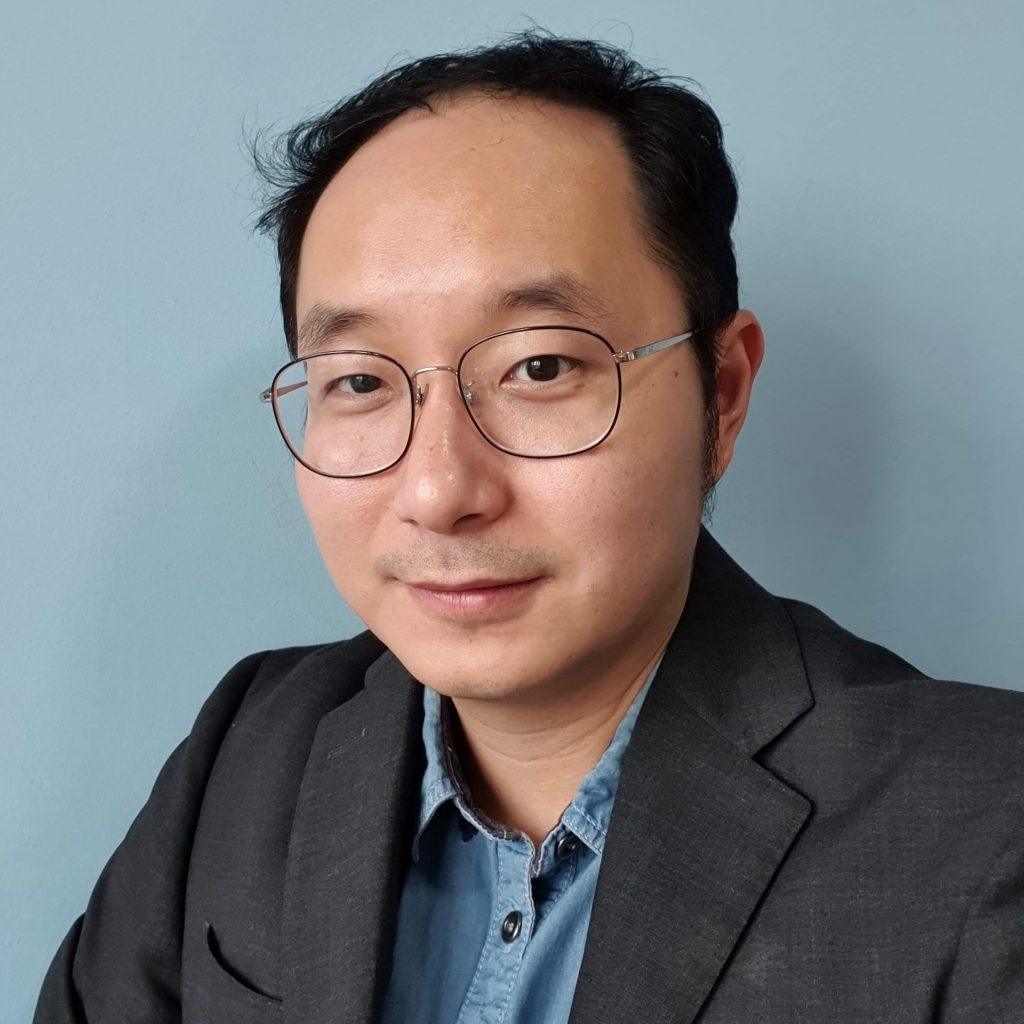}}]
{Bingquan Shen} is currently a Principal Member of Technical Staff in DSO National Laboratories, and an Adjunct Assistant Professor in the Engineering Science Program, National University of Singapore. He received his Honours and PhD in Mechanical Engineering (Control and Mechatronics) from the National University of Singapore in 2009 and 2014 respectively. His research interests lie in intelligent systems, multimodal data understanding/generation, and trustworthy AI.
\end{IEEEbiography}

\vspace{-1.3cm}

\begin{IEEEbiography}[{\includegraphics[width=1in,height=1.25in,clip,keepaspectratio]{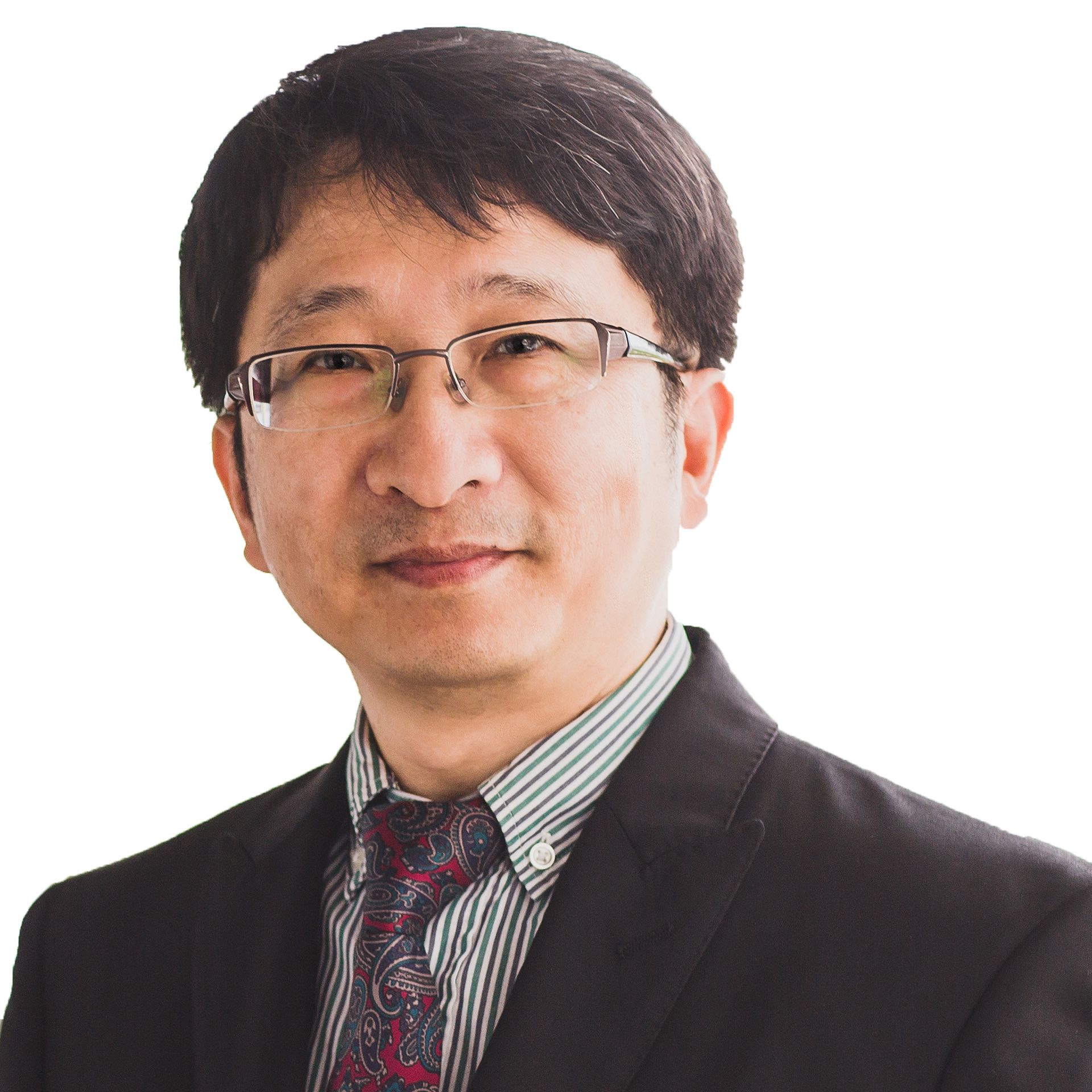}}]{Alex C. Kot} (Life Fellow,  IEEE) has been with the Nanyang Technological University, Singapore since 1991. He was Head of the Division of Information Engineering and Vice Dean Research at the School of Electrical and Electronic Engineering. Subsequently, he served as Associate Dean for College of Engineering for eight years. He is currently Professor and Director of Rapid-Rich Object SEarch (ROSE) Lab and NTU-PKU Joint Research Institute. He has published extensively in the areas of signal processing, biometrics, image forensics and security, and computer vision and machine learning. 

Dr. Kot served as Associate Editor for more than ten journals, mostly for IEEE transactions. He served the IEEE SP Society in various capacities such as the General Co-Chair for the 2004 IEEE International Conference on Image Processing and the Vice-President for the IEEE Signal Processing Society. He received the Best Teacher of the Year Award and is a co-author for several Best Paper Awards including ICPR, IEEE WIFS and IWDW, CVPR Precognition Workshop and VCIP. He was elected as the IEEE Distinguished Lecturer for the Signal Processing Society and the Circuits and Systems Society. He is a Fellow of IEEE, and a Fellow of Academy of Engineering, Singapore.

\end{IEEEbiography}

\vspace{11pt}

\vfill

\end{document}